\documentclass[%
 reprint,
preprintnumbers,
superscriptaddress,
prd,
nofootinbib,
 amsmath,amssymb, aps,
floatfix,
onecolumn
]{revtex4-2}

\usepackage{graphicx}
\usepackage{dcolumn}
\usepackage{bm}
\usepackage{hyperref}
\usepackage{aux_files/aas_macros}
\usepackage{soul}
\usepackage{xspace}
\usepackage{booktabs}

\usepackage[usenames,dvipsnames]{xcolor}
\usepackage{hyperref}
\hypersetup{
    colorlinks = true,
    citecolor = {BrickRed},
    linkcolor = {Blue},
    urlcolor = {Blue}		
}
\usepackage{mathtools}
\usepackage{transparent}
\usepackage{bm}
\usepackage{fontawesome5}
\usepackage{hyperref}

\makeatletter
\newcommand{\github}[1]{%
   \href{#1}{\faGithubSquare}%
}
\makeatother

\newcommand{\Lya}{Ly-$\alpha$\xspace}
\newcommand{\LyaF}{Ly-$\alpha$ forest\xspace}

\newcommand{\abacus}{\textsc{AbacusSummit}\xspace}
\newcommand{\beq}{\begin{equation}}
\newcommand{\eeq}{\end{equation}}
\newcommand{\beqa}{\begin{equation}\begin{aligned}}
\newcommand{\eeqa}{\end{aligned}\end{equation}}

\newcommand{\vk}{\mathbf{k}}

\newcommand{\vq}{\mathbf{q}}
\newcommand{\vp}{\mathbf{p}}
\newcommand{\vx}{\mathbf{x}}
\newcommand{\vs}{\mathbf{s}}

\newcommand{\vpsi}{\boldsymbol{\psi}}

\newcommand{\td}{\delta}
\newcommand{\dd}{\partial}
\newcommand{\kvec}{\mathbf{k}}
\newcommand{\rvec}{\mathbf{r}}

\newcommand{\tr}{\ \mathrm{tr}}

\newcommand{\mat}[1]{\mathbf{#1}}
\newcommand*\diff{\mathop{}\!\mathrm{d}}
\newcommand*\Diff[1]{\mathop{}\!\mathrm{d^#1}}

\newcommand*{\ie}{\emph{i.e.}\xspace}

\newcommand{\knl}{\, k_{\rm NL}\, }

\newcommand{\hinvMpc}{\,h^{-1}\, {\rm Mpc}\xspace}
\newcommand{\hMpcinv}{\,h\, {\rm Mpc}^{-1}\xspace}
\newcommand{\hinvGpc}{\,h^{-1}\, {\rm Gpc}\xspace}

\def\ltsima{$\; \buildrel < \over \sim \;$}
\def\gtsima{$\; \buildrel > \over \sim \;$}
\def\simlt{\lower.5ex\hbox{\ltsima}}
\def\simgt{\lower.5ex\hbox{\gtsima}}

\providecommand{\sorthelp}[1]{}

\providecommand{\sorthelp}[1]{}
\usepackage[dvipsnames]{xcolor}

\newcommand{\BCCP}{Berkeley Center for Cosmological Physics, Department of Physics, UC Berkeley, CA 94720, USA}
\newcommand{\UCBA}{Department of Astronomy, University of California, Berkeley, CA 94720, USA}
\newcommand{\LBL}{Lawrence Berkeley National Laboratory, One Cyclotron Road, Berkeley CA 94720, USA}
\newcommand{\MIT}{Center for Theoretical Physics -- a Leinweber Institute, Massachusetts Institute of Technology, Cambridge, MA 02139, USA}

\newcommand{\pvmhid}[1]{}

\newcolumntype{N}{>{\centering\arraybackslash}m{.5in}}
\newcolumntype{G}{>{\centering\arraybackslash}m{2in}}

\begin{document}

\preprint{MIT-CTP/5942}

\title{The Compressed 3D Lyman-$\alpha$ Forest Bispectrum}

\author{Roger de Belsunce}
        \email{belsunce@mit.edu} 
        \affiliation{\LBL} \affiliation{\BCCP} \affiliation{\MIT}
\author{James M.~Sullivan}
\email{jms3@mit.edu}\thanks{Brinson Prize Fellow} 
\affiliation{\MIT}\affiliation{\BCCP}\affiliation{\UCBA}
\author{Patrick McDonald}
        \affiliation{\LBL}

\begin{abstract}
Cosmological studies of the Lyman-$\alpha$ (\Lya) forest typically
constrain parameters using two-point statistics. However, higher-order statistics, such as the three-point function (or its Fourier counterpart, the bispectrum) offer additional information and help break the degeneracy between the
mean flux and power spectrum amplitude, albeit at a significant computational cost. To address this, we extend an existing highly informative compression of the bispectrum, the skew spectra, to the \Lya forest. We derive the tree-level bispectrum of \Lya forest fluctuations in the framework of effective field theory (EFT) directly in redshift space and validate our methodology on two mock datasets: (i) synthetic 3D \Lya fields using second-order perturbation theory, and (ii) large \Lya forest mocks constructed from the $N$-body simulation suite \textsc{AbacusSummit}. We measure the three-dimensional 
anisotropic cross-spectra between the transmitted flux fraction and all quadratic operators arising
in the bispectrum, yielding a set of 26 skew spectra. 
Using idealized 3D Gaussian smoothing ($R=10\hinvMpc$), we find good agreement ($1$-$2\sigma$ level based on the statistical errors of the mocks) with the theoretical tree-level bispectrum
prediction for monopole and quadrupole up to $k \simlt 0.17 \hMpcinv$. To enable the cosmological analysis of \Lya forest data from the currently observing Dark Energy Spectroscopic Instrument (DESI) and future spectroscopic surveys
, where we cannot do 3D smoothing, we use a line-of-sight smoothing and
introduce a new statistic, the shifted skew spectra. These probe non-squeezed bispectrum triangles and avoid locally applying quadratic operators to the field by displacing one copy of the field in the radial direction. Using a fixed displacement between two points 
of $40\hinvMpc$ (and line-of-sight smoothing $10\hinvMpc$) yields a similar agreement with the theory prediction. For the special case of correlating the squared (and displaced) field with the original one, we analytically forward model the window function making this approach readily applicable to DESI data.
\end{abstract}

\maketitle
\section{Introduction} \label{sec:intro}
The low-density, highly ionized intergalactic medium (IGM) absorbs light, producing a series of absorption features in spectra of distant quasars, called the Lyman-$\alpha$ (\Lya) forest. Along the line-of-sight, it maps small-scale density fluctuations in the neutral hydrogen. Correlating lines of sight yields a large-scale structure (LSS) tracer at high redshifts ($2\leq z \leq 5$) spanning a wide range of scales (for a review, see,~\cite{McQuinn:2016}). 

The quality and quantity of \LyaF data at the present moment is unprecedented and will continue to increase in coming years.
Over the past decades, the number of available medium-resolution spectra from \Lya surveys has dramatically increased from $\sim 210,000$ spectra observed with the extended Baryon Oscillation Spectroscopic Survey (eBOSS; \cite{Dawson:2016, dMdB:2020}) to the currently observing Dark Energy Spectroscopic Instrument (DESI; \cite{DESI:2016, DESI_lya_2024, DESI:2025fxa}) that will observe up to one million spectra over its lifetime. Complementary to those is high-resolution data from the  High Resolution Echelle Spectrometer \citep[HIRES;][]{Vogt:1994,OMeara:2021} and the UV-Visual Echelle Spectrograph \citep[UVES;][]{Dekker:2000,Murphy:2019}, which provide access to smaller scales. 
Future surveys, such as WEAVE-QSO \citep{2016sf2a.conf..259P}, the Prime Focus Spectrograph (PFS; \citep{2022PFSGE}), and 4MOST \citep{2019Msngr.175....3D}, will offer both high- and medium-resolution spectra -- a case in point to develop novel ways of extracting more information from the data at hand.

The spatial statistics of the \LyaF are a powerful probe of cosmological information.
The two-point correlation function (2PCF; \cite{Slosar2013, dMdB:2020}) and the one-dimensional power spectrum along the line-of-sight (P1D; \cite{McDonald06, PYB13, DESI:2023xwh, Karacayli:2023afs}) have been extensively used to test  the Universe’s expansion history via baryon acoustic oscillations \citep[BAO;][]{McDonald:2007,Slosar2013, Busca:2013, dMdB:2020,DESI_lya_2024, DESI:2025zpo} at high redshift ($z=2.33$), as well as the large-scale three-dimensional \Lya correlation function's broadband shape \citep{Slosar2013, Cuceu:2021, Cuceu:2023, Gordon:2023, Cuceu:2025nvl}, neutrino masses \citep{Seljak:2005, Viel:2010, PYB13, Palanque2020}, and primordial black holes \citep{Afshordi:2003, Murgia:2019}. The P1D is particularly sensitive to small-scale physics \citep[see, e.g.,][]{Seljak:2005, Viel:2005,McDonald06, PYB13, Chabanier:2019, Pedersen:2020, 2023MNRAS.526.5118R, 2024MNRAS.tmp..176K}, enabling tests of (warm) dark matter models \citep{Viel:2013, Baur:2016, Irsic17, Kobayashi:2017, Armengaud:2017, Murgia:2018,Garzilli:2019, Irsic:2020, Rogers:2022, Villasenor:2023, Irsic:2023}, early dark energy models \citep{2023PhRvL.131t1001G}, and the thermal properties of the ionized IGM \citep{Zaldarriaga:2002, Meiksin:2009,McQuinn:2016, Viel:2006, Walther:2019, Bolton:2008, Garzilli:2012, Gaikwad:2019, Boera:2019, Gaikwad:2021, Wilson:2022, Villasenor:2022}. Advancements in measuring the three-dimensional power spectrum of the \Lya forest \cite{Font-Ribera:2018, deBelsunce:2024knf, Horowitz:2024, Karacayli:2025tlx}, alternative three-dimensional statistics \cite{Hui_1999, karim2023measurement}, higher-order statistics such as the one-dimensional bispectrum \cite{2003MNRAS.344..776M, delaCruz:2024cai}, and modeling the \Lya forest at the field-level \cite{deBelsunce:2025bqc} motivate the question 
how to extract more, \textit{i.e.}~non-Gaussian, information from the data. 

Analyses of large-scale structure surveys are already extracting higher-order information, either using summary statistics including bispectra within the framework of effective field theory (EFT) of large-scale structure (see, e.g.,~\cite{2022PhRvD.105d3517P, 2024JCAP...05..059D, Chudaykin:2025aux}), marked two-point functions \cite{2016JCAP...11..057W, PhysRevLett.126.011301}, density split statistics \cite{2021MNRAS.505.5731P, 2024arXiv240913583M}, the wavelet scattering transform 
\cite{PhysRevD.109.083535,2022PhRvD.105j3534V}, 
projected bispectra~\cite{Harscouet:2025ksm} or (more ambitiously) directly at the field level \cite{2019A&A...625A..64J}. 
Whilst these approaches have been used in the context of spectroscopic galaxy surveys, higher-order information from the \Lya forest is sparsely exploited and offers a unique opportunity to study the Universe's large-scale structure at high redshifts \cite{2003MNRAS.344..776M, 2004MNRAS.347L..26V, delaCruz:2024cai}. The \Lya forest has four main advantages over traditional galaxy clustering surveys: First, it accesses a much larger cosmological volume which remains inaccessible to galaxy surveys until Stage-V spectroscopy \cite{2022arXiv220903585S}. Second, its high redshift grants access to quasi-linear modes probing fundamental physics, extending the reach of perturbative techniques \cite{Chen:2021, Ivanov2024}. Third, hydrodynamical simulations of the \Lya forest down to $\sim$kpc scales enable precision calibration of perturbative models \cite{McDonald:2001, 2013ApJ...765...39A, Arinyo:2015, 2017MNRAS.464..897B, Givans:2022, Bird:2023evb, deBelsunce:2024rvv, Chudaykin:2025gsh, Abacus_BAO_Lya:2025}. Fourth, shot-noise corrections to the \Lya forest vanish towards higher redshifts (in agreement with the scaling Universe estimate \cite{Ivanov2024, deBelsunce:2024rvv}), offering additional insights compared to galaxies or quasars \cite{Chudaykin:2025gsh}. 
The combination of these factors strongly motivate the development of robust models for higher-order statistics for the \LyaF.

In this work, we extend the skew spectrum approach, which was applied to the 1D \LyaF in \cite{2001ApJ...551...48Z,2003MNRAS.344..776M} and 
galaxies in 
\cite{2015PhRvD..91d3530S, Schmittfull:2021, Dizgah_2020, 2023JCAP...03..045H, 2024JCAP...05..011C, 2024PhRvD.109j3528H}, to the 3D \Lya forest. 
Skew spectra are cross-\textit{power spectra} of quadratic operators applied to the density field (or transmitted flux fraction for the \Lya forest) and the original field. 
They are an efficient proxy statistic for the bispectrum which, similar to the power spectrum, only depends on a single wavenumber with a computational cost only marginally differing from the power spectrum because of pre-processing (weighting) of the quadratic operators applied to the transmitted flux fraction. 
Whilst this approach has been shown to work remarkably well for galaxies \cite{2023JCAP...03..045H, 2024PhRvD.109j3528H}, the lower bias and higher redshift of the \Lya forest will extend the reach (\textit{i.e.}~the $k_{\mathrm{max}}$) of skew spectra and their perturbative prescription. 

For the theory predictions, we derive the tree-level bispectrum of the \Lya forest in the effective field-theory of large-scale structure (EFT). This framework has been successful in describing statistics on large scales only using the relevant symmetries \cite{McDonald:2009,2012JCAP...07..051B, 2012JHEP...09..082C} -- in the present work, directly constructed in redshift space.\footnote{Note that this derivation yields the same terms as when we imagine
anisotropic selection effects in real space which are subsequently transformed to redshift space \cite{Desjacques2018,Ivanov2024}.} Our work is an extension to previous perturbative treatments of the \Lya forest \cite{Seljak:2012, Garny:2018, Givans:2020, Chen:2021, Ivanov2024} and is closely-related to line-of-sight dependent biased tracers \cite{Desjacques2018}. 
Next, we compare the tree-level bispectrum theory predictions to measured skew spectra from two sets of simulations: (i) synthetic \LyaF fields generated using perturbation theory up to second order; and (ii) large-volume \Lya forest mocks painted on top of the $N$-body simulation \abacus \cite{Hadzhiyska:2023, Abacus_BAO_Lya:2025}. 
To ensure applicability of skew spectra to observational data from current and future LSS surveys, we derive and test the exact window function treatment for the simplest of the skew spectra.

The remainder of this paper is organized as follows: 
We briefly review recent existing perturbative models for the \Lya forest and introduce relevant quantities in Sec.~\ref{sec:prelim}, before moving on to an alternative pedagogical derivation in Sec.~\ref{sec:derive_2d}. We introduce the skew spectrum formalism in Sec.~\ref{sec:Lya_skewspectra}. In Sec.~\ref{sec:simulations} we present modified skew spectra that capture non-squeezed configurations of the bispectrum and the results of the numerical tests of our methodology and pipeline on simulations. 
We discuss future work and conclude in Sec.~\ref{sec:conclusions}. 
The Fourier space skew spectra expressions are provided in Appendix~\ref{app:kspace_skew}, 
Appendix~\ref{app:corr_mat} provides the skew spectra multipole correlation matrix,
Appendix~\ref{app:shifted_skew_R5} shows a version of our results in Section~\ref{sec:results_III} for shifted skew spectra with a smaller smoothing scale ($5\hinvMpc$),
we discuss the  window function treatment for a sub-set of skew spectra in Appendix~\ref{app:theory}, and Appendix~\ref{app:disp_ep} and Appendix~\ref{app:consv_Z2} provide details of the derivation of Section~\ref{sec:derive_2d}. 

\section{Preliminaries} \label{sec:prelim}

In this Section, we briefly introduce several quantities that have already appeared in the literature (and their notation) and will arise in the derivation in Section~\ref{sec:derive_2d}.
For further details see, e.g., Refs.~\cite{DJS_review, bias_time_evolution, Chen:2021, Ivanov2024}.

For real-space LSS tracers, the large-scale bias expansion at second order in the density field is
\beq
\label{eq:t_expansion}
\delta_t(\vx) = b_1\delta(\vx) + \frac{b_2}{2}\delta(\vx)^2  + b_{\mathcal{G}_2}\mathcal{G}_2(\vx),
\eeq
where the 2nd-order tidal operator\footnote{We use the Fourier convention 
\begin{equation}
    f(\vk) = \int \Diff3 \vx\, e^{-i\vk\cdot \vx}f(\vx) \quad \leftrightarrow \quad f(\vx) = \int \frac{\Diff3 \vk}{(2\pi)^3}\, e^{i\vk\cdot \vx}f(\vk) = \int_{\vk}\, e^{i\vk\cdot \vx}f(\vk) ,
\end{equation}
with $\left[\nabla_{\vx}f\right](\vk) = i\vk f(\vk)$ and $\int_{\vk}=\int\frac{\diff^3\vk}{(2\pi)^3}$. For the convective derivative we use
$\frac{D}{D\tau}=\dd_{\tau} +v^i \dd_i$. 
Additionally, we use the standard perturbation theory (SPT) kernels \cite{jain:94}
\begin{equation}
F_2(\vk_1, \vk_2) = \frac{5}{7} + \frac{1}{2} \left( \frac{\vk_1 \cdot \vk_2}{k_1 k_2} \right) \left( \frac{k_1}{k_2} + \frac{k_2}{k_1} \right) + \frac{2}{7} \left( \frac{\vk_1 \cdot \vk_2}{k_1 k_2} \right)^2
\,, \quad G_2(\vk_1, \vk_2) = \frac{3}{7} + \frac{1}{2} \left( \frac{\vk_1 \cdot \vk_2}{k_1 k_2} \right) \left( \frac{k_1}{k_2} + \frac{k_2}{k_1} \right) + \frac{4}{7} \left( \frac{\vk_1 \cdot \vk_2}{k_1 k_2} \right)^2\,.
\end{equation}
} is
\begin{equation}
    \mathcal{G}_2[\delta,\delta](\vx)\equiv \left[\frac{\dd_i\dd_j}{\nabla^2}\td(\vx)\right]^2-\td^2(\vx)
\end{equation}
and the related operators $K_{ij}$, $\Pi^{[k]}_{ij}$ are defined as
\begin{align}
& K_{ij}(\vx) = \frac{\dd_i\dd_j\delta(\vx)}{\nabla^2} -\frac{1}{3}\delta_{ij}\delta(\vx)\,,
\end{align}
which is the trace-free part of $\Pi^{\left[1\right]}_{ij}$\footnote{The general expression for the tensor is $\Pi^{\left[n\right]}_{ij} = \frac{1}{(n-1)!}
\left[(\mathcal{H}f)^{-1}\frac{D}{D\tau}\Pi^{[n-1]}_{ij}-\Pi^{[n-1]}_{ij}\right], \quad n>1$ \cite{Desjacques2018,bias_time_evolution}.} 
\begin{align}
& \Pi^{\left[1\right]}_{ij} =\dd_i\dd_j \Phi = \frac{\dd_i\dd_j\delta}{\nabla^2}\,,\\
& \Pi^{[2]}_{ij} = \Pi^{[1]}_{ik} \Pi^{[1]\,k}_j 
  + \frac{10}{21} \frac{\partial_i\partial_j}{\nabla^2} \left(\td^2 - \frac32 K^2 \right) = (K K)_{ij} + \frac23 \td K_{ij} - \frac19 \td^2 \td_{ij}
  + \frac{10}{21} \frac{\partial_i\partial_j}{\nabla^2} \left(\td^2 - \frac32 K^2 \right) \,
\end{align}
with $K^2(\vx) = K_{ij}(\vx)K^{ij}(\vx)$.

For real space tracers that are then displaced into redshift space under the usual transformation from Eulerian position $\vx$ to redshift space position $\vs = \vs^{\rm{RSD}}$
\begin{equation}
    \label{eq:rsd_disp}
    \vpsi_{\mathrm{RSD}} \equiv \vs^{\rm{RSD}}-\vx =  \frac{ v_\parallel}{\mathcal{H}} \hat{z},
\end{equation}
with for some vector $X^i$, $X_\parallel = X^i \hat{z}_i$ leads to the redshift-space density field
\begin{equation}
    \label{eqn:rsd_density}
    \rho(\vs) =\int d\vx~ 
\rho(\vx) \delta^{(D)}(\vs-\vx-
\vpsi_{\mathrm{RSD}}(\vx))
\end{equation}
which, in the plane-parallel approximation with a global line of sight $\hat{z}$, gives the Fourier-space density field (see, e.g.,~\cite{2002PhR...367....1B})
\beq 
\label{eq:rsdmap}
\delta_t^{(s)}(\vk) = \delta_t(\vk) +\int \diff^3x~e^{-i\vk\vx}\left(
e^{-ik_\parallel v_{\parallel}(\vx)/\mathcal{H}}-1\right)(1+\delta_t(\vx))\,.
\eeq

This resulting expression can be expanded perturbatively in the velocity
to the observer's frame 
which, assuming an SPT model for the perturbative velocity expansion (with growth rate $f$), generates the usual RSD kernels \cite{Scoccimarro:1999_rsd,Ivanov2024}
\begin{align}
    \label{eqn:rsd_Zkernels}
    Z_1(\vk;\hat{\mathbf{z}}) &= b_1 +f\frac{k_{\parallel}^2}{k^2}\\
    Z_2(\vk_1,\vk_2;\hat{\mathbf{z}}) &= b_1 F_2(\vk_1,\vk_2) + \frac{b_2}{2} + b_{\mathcal{G}_2}\mathcal{G}_2(\vk_1,\vk_2) + f \frac{k_{3 \parallel}^2}{k_3^2} G_2(\vk_1,\vk_2)\\
    &\quad + \frac{b_1}{2} f \left( \frac{k_{1\parallel}^{2}}{k_1^2} + \frac{k_{2\parallel}^{2}}{k_2^2} \right) +  f^2 \frac{k_{1\parallel}^{2}k_{2\parallel}^{2}}{k_{1}^{2} k_{2}^{2}} \nonumber \\
    &\quad + \frac{b_1}{2} f \left( \frac{k_{1\parallel}k_{2\parallel}}{k_1^2} + \frac{k_{1\parallel}k_{2\parallel}}{k_2^2} \right) + \frac{f^{2}}{2} \frac{k_{1\parallel}k_{2\parallel}}{k_{1}^{2} k_{2}^{2}} \left( k_{1\parallel}^{2} + k_{2\parallel}^{2} \right) \nonumber,
\end{align}
which can be used to model the second order number conserving real-to-redshift-space-displaced tracer density field for in Fourier space.
The fluctuation at second-order is
\begin{equation}
    \delta_F(\vk;\hat{\mathbf{z}}) = Z_1(\vk; \hat{\mathbf{z}})\delta_L(\vk) + \int_{\vk_1,\vk_2} \delta^{(D)}(\vk - \vk_1 -\vk_2) Z_2(\vk_1,\vk_2;\hat{\mathbf{z}})\delta_L(\vk_1)\delta_L(\vk_2)
    \label{eqn:second_order_Fourier},
\end{equation}
where $\delta_L$ is the linear density field.

For the Lyman-$\alpha$ forest, the tracer quantity of interest is the the transmitted Ly$\alpha$ flux fluctuation $\delta^{(s)}_F$.
The bias expansion for $\delta^{(s)}_F$ no longer takes the same form as the expression for real space galaxies that have been displaced into redshift space.
The \Lya forest flux fluctuations only exist in redshift space since (in the top-down type of derivation, e.g., \cite{Seljak2012,Chen:2021,Ivanov2024}) they are due to background quasar rest-frame UV continuum flux photoabsorption along the line of sight due to the optical depth of intervening neutral hydrogen $F=\exp{(-\tau)}$ (see, \cite{McQuinn:2016} for a review). 
Due to the resulting relaxed symmetries of the Lyman-$\alpha$ forest \cite{McDonald:2009,Givans:2020}, further free coefficients must be introduced to produce a consistent bias expansion.
At linear order, this has long been used in the model of the Lyman-$\alpha$ power spectrum \cite{McDonald:1999dt,McDonald:2001} 
\beq 
\td_F = b_1\td + b_{\eta}\eta \,, 
\eeq 
where $\eta = -\dd_{\parallel}v_\parallel/\mathcal{H}$,
is the dimensionless gradient of the peculiar velocity, resulting in the power spectrum
\beq 
P_F(k,\mu) = (b_1 + b_{\eta}f\mu^2)^{2} P_{\rm lin}(k)\,,
\eeq 
where $\mu(\vk,\hat{z})= \frac{k_\parallel}{k}$.
Note that here the sign of $b_\eta$ differs from that in Refs.~\cite{Desjacques2018,Ivanov2024}.

Recently, Ref.~\cite{Ivanov2024} has applied the formalism of Ref.~\cite{Desjacques2018} that was derived to model selection-dependent galaxies (which, without reference to redshift space, do not respect the symmetries of real-space galaxies) to a model of the Lyman-$\alpha$ using the EFT formalism.
Both of these works used the following kernels (which replace the $Z_1,Z_2$ kernels) to model either selection-dependent galaxies or the Lyman-$\alpha$ forest
\begin{align}
    \label{eq:K_kernels}
    K_1(\vk;\hat{\mathbf{z}}) &= b_1 +b_\eta f\frac{k_{\parallel}^2}{k^2}\\
    K_2(\vk_1,\vk_2;\hat{\mathbf{z}}) &= b_1 F_2(\vk_1,\vk_2) + \frac{b_2}{2} + b_{\mathcal{G}_2}\mathcal{G}_2(\vk_1,\vk_2) + b_\eta f \frac{k_{3 \parallel}^2}{k_3^2} G_2(\vk_1,\vk_2) \\
    &\quad + \frac{b_{\delta \eta}}{2} f \left( \frac{k_{1\parallel}^{2}}{k_1^2} + \frac{k_{2\parallel}^{2}}{k_2^2} \right)  + b_{\eta^2} f^2 \frac{k_{1\parallel}^{2}k_{2\parallel}^{2}}{k_{1}^{2} k_{2}^{2}} \nonumber \\
    &\quad + \frac{b_1}{2} f \left( \frac{k_{1\parallel}k_{2\parallel}}{k_1^2} + \frac{k_{1\parallel}k_{2\parallel}}{k_2^2} \right) + b_{\eta} \frac{f^{2}}{2} \frac{k_{1\parallel}k_{2\parallel}}{k_{1}^{2} k_{2}^{2}} \left( k_{1\parallel}^{2} + k_{2\parallel}^{2} \right) \nonumber \\
    &\quad+ b_{(KK)_\parallel} (KK)_{\parallel}(\vk_1,\vk_2;\hat{\mathbf{z}}) + b_{\Pi^{[2]}_\parallel} \Pi^{[2]}_{\parallel}(\vk_1,\vk_2;\hat{\mathbf{z}}) \nonumber
\end{align}
where we adopt the notation of \cite{Schmittfull:2021}\footnote{We will keep this notation to make comparison with previous work on skew spectra easy, but in this work, it has the unfortunate consequence that a field or wavevector with a $_\parallel$ subscript contains a \textit{single} factor of $\hat{z}$, while a $^\parallel$ superscript contains \textit{two} factors of $\hat{z}$.} for the quadratic operators and explicitly write out the angle $\mu_i^2 \equiv \frac{k_{i\parallel}^{2}}{k_{i}^{2}}$,
and where the new operators are defined as
\begin{align} \label{eq:KKpar}
    (KK)_{\parallel}(\vk_1,\vk_2;\hat{\mathbf{z}}) &= K_{ij} K_{jl} \hat{z}^i\hat{z}^l\\
    &=\frac{k_{1\parallel}k_{2\parallel}}{k_{1}^{2}k_{2}^{2}}\left(\vk_1 \cdot \vk_2 \right) - \frac13 \left(\frac{k_{1\parallel}^{2}}{k_{1}^{2}} + \frac{k_{2\parallel}^{2}}{k_{2}^{2}}\right) + \frac19,
    \nonumber
\end{align}
and
\begin{align} \label{eq:Pi22par}
    \Pi^{[2]}_{\parallel}(\vk_1,\vk_2;\hat{\mathbf{z}}) &= \Pi^{[2]}_{ij}\hat{z}^i\hat{z}^j\\
    &= \frac{k_{1\parallel}k_{2\parallel}}{k_{1}^{2}k_{2}^{2}}\left(\vk_1 \cdot \vk_2 \right) + \frac57 \frac{k_{3\parallel}^{2}}{k_{3}^{2}}\left(1 - \frac{\left(\vk_1 \cdot \vk_2 \right)^{2}}{k_{1}^{2}k_{2}^{2}}\right),
    \nonumber
\end{align}
with $\vk_{3} = -\vk_1 - \vk_2$.
This results in the quadratic bias expansion
\beq
\label{eq:F_full}
\begin{split}
\delta_F(\vs) = b_1\delta(\vx) &+ \frac{b_2}{2}\delta^2(\vx)  + b_{\mathcal{G}_2}\mathcal{G}_2(\vx)\\
&+ b_\eta \eta(\vx)  + b_{\delta\eta}[\eta \delta](\vx) + b_{\eta^2}\eta^2(\vx)\\
&+ b_{(KK)_\parallel} KK_\parallel(\vx) + b_{\Pi^{[2]}_\parallel} \Pi^{[2]}_{\parallel}(\vx).
\end{split}
\eeq

In the next Section (\ref{sec:derive_2d}), we derive the Lyman-$\alpha$ flux overdensity bias expansion by appealing directly to the symmetries of the Lyman-$\alpha$ forest \cite{McDonald:2009, Givans:2020, Ivanov2024,Desjacques2018,Chen:2021}.
The only novelty of this next Section 
is in the conceptual development of the expansion of $\delta_{F}(\vs)$ without reference to these selection effect operators 
- we do not claim to develop any new technical results.

\section{EFT bias expansion without line-of-sight symmetry}
\label{sec:derive_2d}
We now provide a derivation of the Lyman-$\alpha$ forest bias expansion by referring only to the symmetries of the system.
The \Lya forest is fundamentally a redshift-space phenomenon, and we should be able to describe it 
without reference to ``rest-frame'' quantities, as it does not even exist in real space.
In the bottom-up perturbative construction, there is no reason that the \Lya forest flux fluctuation overdensity field, $\delta_F$, should have anything to do with the standard redshift space transformation that is applied to the galaxy overdensity, as, for flux, there is no sense of 
number conservation 
that is assumed for galaxies (or neutral hydrogen gas).
By building the bias expansion only with the symmetries actually obeyed by the \Lya flux overdensity field in the plane-parallel approximation (translation invariance, $SO(2)$ rotation invariance about the line of sight, sightline reflection symmetry, and the equivalence principle. Sec.~\ref{subsec:symm}), we construct a general and pedagogical perturbative biasing description of the forest (Sec.~\ref{app_sub:bias_expand}) with the \textit{SO}(2) basis.
This requires us to write down the appropriate perturbative expansion terms (Sec.~\ref{subsubsec:det_bias_pt}), address restrictions required by the equivalence principle (Sec.~\ref{subsubsec:ep}), and enumerate the higher-derivative terms (Sec.~\ref{app_sub:derivative}) and stochastic fields (Sec.~\ref{app_sub:stoch}).
We close this pedagogical section by showing how to reduce the $SO(2)$ expansion to the biasing description of a tracer field that respects the usual set of 3D symmetries after it undergoes the usual real-to-redshift-space transformation, which is relevant for recovering the case of redshift-space galaxies (Sec.~\ref{app_sub:restore_los}).
At all times in this work, we employ the plane-parallel approximation with global line of sight vector $\hat{z}$.
We assume adiabatic, Gaussian initial fluctuations.

\subsection{Symmetries of the \LyaF \label{subsec:symm}}
A LSS tracer in 3D respects large-scale homogeneity and isotropy.
A tracer $\delta_t$ respecting large-scale homogeneity has two-point correlation function
\begin{equation}
    \label{eqn:homogeneity}
    \langle \delta_t(\vs)\delta_t(\vs')\rangle = \langle \delta_t(\vs+\Delta)\delta_t(\vs'+\Delta)\rangle = \xi_{tt}(\vs-\vs'),
\end{equation}
i.e. there is no preferred location in space - the correlator is translation invariant.\footnote{We will work at the level of two-point functions, since that is simplest, but of course the usual logic extends to higher-point functions.}

The presence of a preferred line of sight breaks isotropy for rotations through the $x-y$ (transverse) plane.
However, the correlator is still invariant to rotations inside the plane ($R_{xy}$):
\begin{equation}
    \label{eqn:2d_isotropy}
    \langle \delta_t(\vs)\delta_t(\vs')\rangle = \langle \delta_t(R_{xy}\vs)\delta_t(R_{xy}\vs')\rangle = \xi_{tt}(|\vs_\perp-\vs_\perp'|,\vs_\parallel-\vs_\parallel'),
\end{equation}
i.e. there are no preferred directions in the transverse plane.

The form of the correlator in eqn.~\ref{eqn:2d_isotropy} still allows for a preferred direction even in the absence of rotations through the $x-y$ plane.
Specifically it allows for a dependence on the sign of $\vs_\parallel-\vs_\parallel'$.
Clearly this is unsatisfactory, as there is no physical reason for the density field to discriminate between displacements oriented away from the observer or towards the observer, even when rotations through the transverse plane are treated differently from those inside the transverse plane.
We therefore also require this ``sightline'' symmetry to remove the unphysical sign dependence.
An alternative way of viewing this symmetry is as what might be called ``1D isotropy'' of the tracer field for all values of $\vs_\parallel$\footnote{In reality, it is the manifestation of the only non-trivial 1D symmetry group. This group is intuitively similar to an aspect of the group $D_{\infty h}$ (which is well-known in chemistry as the symmetry group of homonucleonic diatomic molecules) used in App. B of Ref.~\cite{Givans:2020} to characterize \Lya flux for a system of a single neutral hydrogen cloud once one considers replicating this group structure at every point along the line of sight.}.
This precludes odd factors of $\hat{z}$, which would mean there is a way to distinguish the direction toward the observer along the line of sight from the direction away from the observer along the line of sight at a given value of $\vs_\parallel$.

With the addition of this symmetry, the $SO(2)$ tracer correlator becomes
\begin{equation}
    \label{eqn:1d_isotropy}
    \langle \delta_t(\vs)\delta_t(\vs')\rangle = \langle \delta_t(\vs\odot\mathrm{sign}[s_{\parallel}\hat{z}])\delta_t(\vs'\odot\mathrm{sign}[s_{\parallel}'\hat{z}])\rangle = \xi_{tt}(|\vs_\perp-\vs_\perp'|, |\vs_\parallel-\vs_\parallel'|),
\end{equation}
where $\odot$ is the elementwise product.
These symmetries result in two point functions that can be described in terms of $(\vs_\perp, \vs_\parallel)$ or $(s, \hat{\vs}\cdot \hat{z})$
and power spectra that can be described (in the usual manner) in terms of $(\vk_\perp, \vk_\parallel)$ or $(k,\mu_z=\hat{\vk}\cdot \hat{z})$

With these symmetries in mind, we now proceed to write down the bias expansion consistent with these symmetries.
We return to the application of the equivalence principle in Section~\ref{subsubsec:ep}.

\subsection{\label{app_sub:bias_expand} The \Lya flux overdensity bias expansion}

A tracer that respects the usual set of 3D symmetries is modeled by expanding its overdensity using a bias expansion that reflects these symmetries.
The expansion basis of this tracer field includes all fields up to a certain perturbative order that can be constructed out of locally observable combinations of the (Newtonian) gravitational potential.
The amplitude of these coefficients are then interpreted as Lagrangian or Eulerian bias parameters, where the former are related to the latter through a perturbative forward model for the matter density field and related observables.
When considering these tracers in redshift-space, the bias expansion must be modified to allow for (a restricted set of) additional line-of-sight dependent basis fields with prescribed amplitudes.
In contrast, for an $SO(2)$ tracer, the basis of the expansion must include line-of-sight-dependent fields from the outset, and these fields are accompanied by free coefficients.

Just as for the Lagrangian and Eulerian pictures of the bias expansion, there are multiple ways to interpret this fact about the $SO(2)$ expansion.
One can imagine generating the basis fields for an $SO(2)$ tracer, like the \Lya forest, by applying a displacement $\vpsi_{\mathrm{RSD}}(\vx)$ to basis fields taking values in Eulerian space $\mathcal{O}(\vx)$, and well as considering ``selection-dependent'' operators with free coefficients as an additional step.
This is effectively the top-down construction, e.g. as it relates to the FGPA\footnote{The FGPA is an analytic approximation assuming an equilibrium between optically-thin photoionization and collisional recombination of residual neutral hydrogen in the IGM%
. It yields a power-law relation between the density field, $\td$, and the resulting optical depth $\tau \propto (1+\td)^{\beta}$  \cite{1997MNRAS.292...27H, 1998ApJ...495...44C}.}.
This approach is most natural when discussing the standard picture of biased tracers that form depending on the 3D symmetric basis fields, such as galaxies, that undergo redshift-space distortions.
However 
this approach requires invoking selection dependence in \textit{real space} as an additional property of the tracer field, which is a not a natural way to interpret the \LyaF.

Alternatively, one can think of $SO(2)$  tracers like the \Lya forest as depending on ``Lagrangian-$SO(2)$'' basis fields in a formation region of Lagrangian-$SO(2)$ space.
A hypothetical Lagrangian observer would see this Lagrangian-$SO(2)$ \Lya forest even in the absence of gravitationally-generated non-linearity\footnote{This, of course, applies to the case of non-Ly$\alpha$ tracers, like galaxies in redshift space, as well}.
To connect to low-redshift observations, the thus obtained Lagrangian line-of-sight-dependent bias expansion is then modified by gravitational nonlinearity modeled perturbatively in the usual way.
The various ways of viewing the same final result can be considered as a $SO(2)$ augmentation of the ``monkey bias'' picture Ref.~\cite{Fujita:2020_monkey_bias}, which we touch on again in Appendix~\ref{app:disp_ep}.

We now write down all possible terms allowed in the $SO(2)$ bias expansion, explicitly remarking on which terms are allowed by the symmetries.

\subsubsection{\label{subsubsec:det_bias_pt} Perturbative expansion}

To construct the basis of bias operators used to model a field $\delta_{F}(\vs)$ that may depend on quantities along the line of sight, we consider all physical operators that can be constructed from contractions of $\hat{z}^{l}\hat{z}^{m}\partial_{i}\partial_{j}\Phi(\vs)$ up to second order\footnote{Here $\Phi$ is normalized such that $\partial^{2}\Phi = \delta$},
where derivatives are defined with respect to $\vs$.
This is consistent with the assumption that any field of interest $\delta_F(\vs)$ may only depend on operators consistent with the equivalence principle (second and higher derivatives of $\Phi(\vs)$).
We neglect all operators related to the baryon and dark matter relative perturbations \cite{DJS_review,Givans:2020}.

Without making any reference to rest-frame quantities, we can enumerate all the operators at a given order in $\partial_i \partial_j \Phi$ by contracting all possible indexed quantities, including the line of sight.
At linear order this produces $\delta(\vs)$ and $\delta^{\parallel}(\vs)$ (via $\tr\left( \Pi_{ij}^{[1]}|_{(1)}\right)(\vs)$ and $\hat{z}^{i}\hat{z}^{j} \Pi_{ij}^{[1]}|_{(1)}(\vs)$).
Since $\delta(\vs)$ is an operator with the ``full symmetry'' of 3D rotations is obviously a subset of all operators that are symmetric under the symmetry group about the line of sight.
Also, we use the notation $\delta^{\parallel} \equiv \Pi^{[1]}|_{(1)}^{\parallel}$ to 
emphasize that we need not introduce a velocity field explicitly to construct the bias operators.

At second order, we also obtain the operators $\delta^{2}(\vs)$, $K^{2}(\vs)$, $\delta(\vs)\delta^{\parallel}(\vs)$, $(\delta^{\parallel}(\vs))^{2}$, $\Pi^{[2]}|_{(2)}^{\parallel}(\vs)$, $(KK)^{\parallel}(\vs)$.
The fact that these operators constitute all possible contractions of line-of-sight vectors and factors of $\partial_i \partial_j \Phi(\vs)$  
can be shown by writing down two columns each with 4 entries, one with $\hat{z}^{i},\hat{z}^{j},(\hat{z}^{l},\hat{z}^{m})$ and one with $\partial_i,\partial_j,(\partial_l,\partial_m)$
and drawing all possible non-crossing, edge-length 1 graphs between these elements that have 2 (4) edges in the case of 2 (4) index contractions. 
This generates all terms except for those arising from higher-order part of the potential ($\Phi = \Phi^{(1)}+\Phi^{(2)}+...$), this includes those obtained from e.g. the parallel displacement of $\delta^2$, or, $(\delta^{2})^\parallel$ (See also Appendix C of Ref.~\cite{Chen:2021}.).
Upon adding these, we exhaust all possible contributions at 2nd order, and the deterministic bias expansion for our line-of-sight-dependent tracer is then:
\beq
\label{eq:app_F_2d}
\begin{split}
\delta_F(\vs) &= b_1\delta(\vs) + b_\parallel \delta^{\parallel}(\vs)\\
&+ \frac{b_2}{2}\delta^2(\vs)  + b_{\mathcal{G}_2}\mathcal{G}_2(\vs)+ b_{(KK)_\parallel} (KK)_{\parallel}(\vs)+ b_{\delta\parallel}\delta(\vs) \delta^{\parallel}(\vs) + b_{\parallel^2}(\delta^{\parallel})^2 +b_{\Pi^{[2]}_\parallel}\Pi^{[2]}|_{2}^{\parallel}(\vs)  \\
&\quad 
+ \delta_F^{(\rm HD)}(\vs) + \epsilon^{(\rm{tot})}(\vs),
\end{split}
\eeq
where the higher-derivative and stochastic terms on the last line are given in Sec.~\ref{app_sub:derivative} and Sec.~\ref{app_sub:stoch}, respectively.
We have neglected terms of the form $\hat{z}^l\hat{z}^m\partial_i\Phi \partial_j \delta$, and discuss such displacement terms in Section~\ref{subsubsec:ep}.

\subsubsection{\label{subsubsec:ep} Equivalence principle}


In Section~\ref{subsubsec:det_bias_pt} we neglected displacement terms of the form $\{b_{\psi\nabla \delta} ~\vpsi(\vs) \nabla \delta(\vs),b_{\psi\nabla \delta} ~\vpsi(\vs) \nabla \delta^\parallel(\vs)\}$ and\\  $\{b_{\psi_\parallel \partial_\parallel \delta^\parallel} ~\psi_\parallel(\vs) \partial_\parallel \delta(\vs),b_{\psi_\parallel \partial_\parallel \delta^\parallel} ~\psi_\parallel(\vs) \partial_\parallel \delta^\parallel(\vs)\}$, all of which are second order in the density field for a linear displacement and satisfy the previously-stated symmetries (SO(2), LOS isotropy, translations).
This was implicit in our assumption that all relevant operators can be constructed from $\partial_i\partial_j\Phi(\vs)$ only, rather than from any terms involving $\partial_i \Phi$, which are explicitly assumed not to contribute e.g. in Refs.~\cite{McDonald:2009,bias_time_evolution} by neglecting a homogeneous velocity or assuming tracers move on DM fluid trajectories\footnote{Of course, terms involving only powers of $\Phi$ are also forbidden by the equivalence principle as unobservable for Gaussian initial conditions \cite{DJS_review}, though see \cite{2025:Sullivan_lpngte} for a potentially useful interpretation of this unobservable influence on tracer formation.}.

From a mathematical perspective, this is not the most general combination of operators generated by the equations of motion under consideration, even in the full 3D symmetric case.
However, upon invoking the equivalence principle, assuming gravitational interactions in GR for adiabatic Gaussian initial density fluctuations, it is possible to fix the bias parameters using the LSS consistency relations \cite{Fujita:2020_monkey_bias,kehagias_consistency_2,kehagias_riotto_consistency_1,creminelli_ep_1,horn_consistency,Peloso:2013zw}.
E.g. for the case of real-space tracers, Ref.~\cite{Fujita:2020_monkey_bias} finds 
\begin{equation}
    \label{eqn:cr_3d}
    \lim_{q\to0}\frac{\langle \delta(\vq) \delta_{F}^{(a)}(\vk_1) \delta_{F}^{(b)}(\vk_2)\rangle'}{P_m(q)\langle\delta_{F}^{(a)}(\vk_1)\delta_{F}^{(b)})(\vk_2)\rangle'} = \frac{ \vq \cdot \vk_1}{q^2} \left(b_1^{(a)}b_{\psi\nabla \delta}^{(b)}  - b_1^{(b)} b_{\psi\nabla \delta}^{(a)}\right),
\end{equation}
which implies that terms of the form $b_{\psi\nabla \delta} ~\vpsi(\vs) \nabla \delta(\vs)$, e.g. as arises in the $F_2$ term, cannot enter the bias expansion with a free coefficient  $b_{\psi\nabla \delta}$, and this coefficient must take the value $b_{\psi\nabla \delta} = b_1$ \cite{Fujita:2020_monkey_bias,DAmico:2021_lss_bootstrap}\footnote{
This can be seen
by simply considering the soft limit of the matter-matter-tracer power spectrum $\lim_{q\to0} \frac{\langle \delta(\vq)\delta(\vk_1)\delta_F^{(a)}(\vk_2)\rangle'}{P(q)\langle\delta(\vk_1)\delta(\vk_2)\rangle'} \propto \left( b_1^{(a)} - b_{\psi\nabla\delta}^{(a)}\right) \to 0$}\footnote{This real-space relationship is also accounted for in redshift-space versions of the expressions below.}.

In a similar manner, we extend this logic to the case of an $SO(2)$ tracer.
Due to the adiabatic mode condition, without necessarily assuming anything about displacing real space galaxies into redshift space (e.g. not assuming density conservation from a real-to-redshift space transformation), we can similarly consider the squeezed limit of the $SO(2)$ tracer bispectrum with a soft matter mode.
This takes almost the same form as the galaxy redshift-space consistency relation \cite{creminelli_ep_2,kehagias_consistency_2}.

We now extend this line of argument for isotropic displacements to the case of the line-of-sight displacement-type terms $b_{\psi_\parallel \partial_\parallel \delta} ~\psi_\parallel(\vs) \partial_\parallel \delta(\vs)$.
Again, the suitably $SO(2)$-generalized consistency relation removes both the free coefficients that could possibly arise at 2nd order (assuming a unique linear displacement)
\begin{equation}
    \label{eqn:cr_3d_rsd_1}
    \lim_{q\to0}\frac{\langle \delta(\vq) \delta_{F}^{(a)}(\vk_1) \delta_{F}^{(b)}(\vk_2)\rangle'}{P_m^s(q)\langle\delta_{F}^{(a)}(\vk_1)\delta_{F}^{(b)}(\vk_2)\rangle'} \supset \frac{ k_1 ~p(\mu_q,\mu_{1z})}{q} \left(b_\parallel^{(a)}b_{\psi_\parallel \partial_\parallel \delta^\parallel}^{(b)}  - b_\parallel^{(b)} b_{\psi_\parallel \partial_\parallel \delta^\parallel}^{(a)}\right),
\end{equation}
and 
\begin{equation}
    \label{eqn:cr_3d_rsd_2}
    \lim_{q\to0}\frac{\langle \delta(\vq) \delta_{F}^{(a)}(\vk_1) \delta_{F}^{(b)}(\vk_2)\rangle'}{P_m^s(q)\langle\delta_{F}^{(a)}(\vk_1)\delta_{F}^{(b)}(\vk_2)\rangle'} \supset \frac{ k_1 ~p(\mu_q, \mu_{1z})}{q} \left(b_1^{(a)}b_{\psi_\parallel \partial_\parallel \delta}^{(b)}  - b_1^{(b)} b_{\psi_\parallel \partial_\parallel \delta}^{(a)}\right),
\end{equation}
where $p(\mu_{1z})$ is a polynomial in $\mu_{1z}, \mu_q$.
These two terms correspond to the IR-divergent terms on line 3 of eqn.~\ref{eq:K_kernels} when mode in the denominator becomes long.

It is also easily seen from a similar calculation involving the $G_2^{\parallel}$ term that the 3D consistency relation also precludes any terms of the form 
$b_{\psi\nabla \delta^\parallel} ~\vpsi(\vs) \nabla \delta^\parallel(\vs)$, again, which are restricted to have no free coefficient, but rather $b_{\psi\nabla \delta^\parallel}=b_\eta$ 
\begin{equation}
    \label{eqn:cr_3d_vel}
    \lim_{q\to0}\frac{\langle \delta(\vq) \delta_{F}^{(a)}(\vk_1) \delta_{F}^{(b)}(\vk_2)\rangle'}{P_m^s(q)\langle\delta_{F}^{(a)}(\vk_1)\delta_{F}^{(b)}(\vk_2)\rangle'} \supset \frac{ k_1 ~p(\mu_q,\mu_{1z})}{q} \left(b_\parallel^{(a)}b_{\psi \nabla \delta^\parallel}^{(b)}  - b_\parallel^{(b)} b_{\psi \nabla \delta^\parallel}^{(a)}\right)
\end{equation}
This list includes all IR-divergent terms at 2nd order - the other terms, including those related to $(KK)_\parallel$ and $\Pi_\parallel^{[2]}$, do not contain such divergences.
We emphasize that nowhere did we make reference to the real-to-redshift space transformation to derive the fixing of these bias parameters, only appealing to the symmetries of the system of $SO(2)$ tracers and the adiabatic mode condition.

We supply a more detailed calculation in Appendix~\ref{app:disp_ep}.
There we describe that, in fact, the $SO(2)$ bias parameters can be fixed exactly beyond the relations described above by simply considering the matter-matter-tracer cross spectra.

\subsubsection{\label{app_sub:derivative} Higher-derivative contributions}

We can obtain all necessary higher-derivative contributions at linear order in the perturbations through the usual strategy of performing a semi-local Taylor expansion \cite{McDonald:2009,DJS_review} and integrating over an unknown kernel that obeys the symmetries of the system\footnote{As with the rest of this section, we work directly at the level of the observed position $\vs$, though at linear order this is no different from any other position variable (i.e. real space $\vx$).}.
Unlike in the usual isotropic treatment, we crucially allow the kernel to be anisotropic due to the line-of-sight dependence:
\begin{equation}
    \label{eq:appa_kernel_aniso}
    K(|\vs'-\vs|) \to K(\vs,\vs') = K^{(\mathrm{aniso})}(|\vs-\vs'|,|\vs_{\parallel}-\vs_{\parallel}'|)
\end{equation}
where $K^{(\mathrm{aniso})}$ is even in its arguments due to the assumed symmetry along the line of sight (see the discussion in Sec.~\ref{subsec:symm}).
Otherwise, the resulting anisotropic kernel is fully general.

By expanding both $SO(2)$ linear perturbations $\delta(\vs)$ and $\delta^\parallel(\vs)$ (with kernels $K_\delta^{(\rm{aniso})}$ and $K_{\delta^\parallel}^{(\rm{aniso})}$, respectively), we generate the leading higher-derivative contributions in the line-of-sight anisotropy setting:
\begin{align}
    \label{eq:appa_higher_deriv}
    \int_{\vs'} K^{(\mathrm{aniso})}_{\delta}(|\Delta\vs|,|\Delta\vs_{\parallel}|)\delta(\vs') & = \int_{\Delta \vs} K_{\delta}^{(\mathrm{aniso})}(|\Delta\vs|,|\Delta\vs_{\parallel}|)\big[\delta(\vs)
    + \partial_{i} \delta(\vs) \Delta\vs^{j} 
    + \frac12 \partial_i\partial_j\delta(\vs)\Delta\vs^{l} \Delta\vs^{m} + ...\big], \\
    \int_{\vs'} K^{(\mathrm{aniso})}_{\delta^\parallel}(|\Delta\vs|,|\Delta\vs_{\parallel}|) \delta^\parallel(\vs')  &= \int_{\Delta \vs} K^{(\mathrm{aniso})}_{\delta^\parallel}(|\Delta\vs|,|\Delta\vs_{\parallel}|)\big[ \delta^\parallel(\vs)
    + \partial_{i} \delta^\parallel(\vs) \Delta\vs^{j}
    + \frac12 \partial_i\partial_j\delta^\parallel(\vs)\Delta\vs^{l} \Delta\vs^{m} + ...\big],
\end{align}
with $\Delta\vs(\vs,\vs') \equiv \vs-\vs'$,  $\Delta\vs_{\parallel}(\vs,\vs') \equiv \vs_{\parallel}-\vs_{\parallel}'$.
As in the isotropic case, we interpret $b_1 = \int_{\Delta \vs}K_\delta^{(\rm{aniso})}(|\Delta \vs|,|\Delta \vs_{\parallel}|)$, and, similarly, interpret $b_{\delta^\parallel} = \int_{\Delta \vs}K_{\delta^\parallel}^{(\rm{aniso})}(|\Delta \vs|,|\Delta \vs_{\parallel}|)$.
The first line generates the standard terms in the isotropic case proportional to $\nabla^{2}\delta$ as well as a linear gradient term that we neglect by the isotropy of $K^{(\mathrm{iso})}$ \cite{McDonald:2009}.
This line \textit{also} now generates a new term from the anisotropy of the kernel proportional to $\partial_{\parallel}^{2} \delta$, as well as a linear term from the gradient along the line of sight of the form $\partial_{\parallel} \delta$, which we neglect based on the line-of-sight symmetry discussed in Section~\ref{subsec:symm}.
The second line generates analogous 2nd-order in derivative terms for $\delta^\parallel$
The higher-derivative contributions can then be written, at leading order in perturbations, as
\begin{align} \label{eqn:bias_hd}
\delta_t^{(\mathrm{HD})}(\vs) &= b_{\nabla^{2}\delta} \nabla^{2}\delta(\vs) + b_{\nabla^{2}\delta^\parallel} 
 \nabla^{2}\delta^\parallel(\vs) \nonumber \\
&\quad+ b_{\partial_\parallel^2 \delta} \partial_{\parallel}^2 \delta(\vs) + b_{\partial_\parallel^2 \delta^\parallel} \partial_{\parallel}^{2} \delta^{\parallel}(\vs).
\end{align}
While the second and third terms have the same $\mu^2 
k^2$ dependence in Fourier space, in principle they must have independent bias parameters.
We note that, up to this last point, these are the same terms presented in Ref~\cite{Desjacques2018,Ivanov2024}, and are degenerate with the velocity biases presented there, though here we have no need to explicitly discuss the velocity field.
All perturbative counterterms that would be generated in the perturbative description of the RSD mapping at the order that we work at \cite{Ivanov2024,Desjacques2018,senatore_zaldarriaga_rsd_eft} are already degenerate with the terms we include that are required by $SO(2)$ symmetry\footnote{This illustrates the conceptual clarity of the $SO(2)$ approach to the bias expansion, as one cannot leave out terms that must then be restored by considering counterterms due to loops in the perturbative RSD treatment, a problem that was highlighted in the top-down approach of Ref.~\cite{Ivanov2024}.}, so we do not discuss them explicitly. 
For the \LyaF, the size of these terms may not have the usual association with the formation scale $R_\star$ usually invoked for halos, e.g., due to large-scale gradients from fluctuating backgrounds \cite{Cabass2019:uvb_grad,GontchoAGontcho2014:bg_fluc_deriv}.

\subsubsection{\label{app_sub:stoch} Stochastic terms}
Stochastic fields arise due to small-scale fluctuations and are by construction uncorrelated with (large-scale) biasing operators.
For the most general $SO(2)$ tracer, the tracer stochasticity can be expressed, at leading order in perturbations, and at 2nd
order in derivatives as
\begin{align}
    \label{eq:add_stoch_0}
    \epsilon^{(\rm{tot})}(\vs) &= \tilde{\epsilon}^{(\rm s)}(\vs)(1 + k^2 + k^i \hat{z}_i + (k^{i} \hat{z}_i)^2 )  \nonumber \\
    &\quad + \tilde{\epsilon}_{(\rm v)}^{i}(\vs)(k_i + \hat{z}_i + \hat{z}^j k_j \hat{z}_i + \hat{z}_j k_i k^j + k^2 \hat{z}_i)  \nonumber \\
    &\quad +  \tilde{\epsilon}_{(\rm t)}^{ij}(\vs)(k_i k_j + \hat{z}_i \hat{z}_j + 2 k_i \hat{z}_j + \hat{z}^k \hat{z}_i k_k k_j + \hat{z}_i \hat{z}_j k^2 + k_k k_l \hat{z}^k\hat{z}^l \hat{z}_i\hat{z}_j )
\end{align}
where we have separated possible contributions from up to second rank tensors up to second order in derivatives (or $k$) and considering up to 4 factors of the line of sight $\hat{z}$.
Unlike the usual case of the real-space (3D symmetry) fields, we now allow for vector and tensor perturbations to be sourced by the line-of-sight vector\footnote{In the case of real-space galaxies that are displaced into redshift space, the only physical source of stochasticity can be due to the line-of-sight velocity. As a result, in this case, the tensor stochasticity should be thought of as the product of two vector stochastic fields. We thank Fabian Schmidt for pointing this out.}.
As in the usual case, we obtain the second-order terms of the form $\epsilon_\delta \delta$\footnote{We neglect higher derivatives of these terms under the assumption that the (appropriately anisotropic) nonlocality scale $R^a_\star$ contributes with a similar size as the perturbative nonlinear scale and so is less important in the large-scale limit than the other terms due to suppression by both the nonlocality scale and higher perturbative order.}.
Here all terms are effective stochastic fields, and contain contributions from contractions of vector and tensor stochasticities.

Collecting terms in Eq.~\eqref{eq:add_stoch_0} into effective stochasticities, and neglecting terms with ``bare'' factors of $\hat{z}$ by line-of-sight isotropy (Sec.~\ref{subsec:symm}), we obtain 

\begin{align} \label{eq:add_stoch_1}
    \epsilon^{(\rm{tot})}(\vs) &= \tilde{\epsilon}^{(\rm s)}(\vs)(1 + k^2 + k_\parallel^2 )  \nonumber \\
    &\quad + \tilde{\epsilon}_{(\rm v)}^{i}(\vs)(k_i + k_\parallel \hat{z}_i )  \nonumber \\
    &\quad +  \tilde{\epsilon}_{(\rm t)}^{ij}(\vs)(k_i k_j + k_\parallel \hat{z}_i k_j  ),
\end{align}
where each $\epsilon_{a}(\vs)$ is now an effective stochasticity, absorbing multiple terms from Eq.~\eqref{eq:add_stoch_0} (specifically between the tensors and scalars).

There are two vectors with one index in the $SO(2)$ symmetry, $k^i$ and $\hat{z}^i$.
So, if it is further assumed that  $\epsilon_{(\rm v)}^{i}(\vs) \propto k^i$ and $\epsilon_{(\rm t)}^{ij}(\vs) \propto k^i k^j$ (as assumed in \cite{Perko:2016}), at second order in derivatives, the tensor terms are higher order ($k^4$)\footnote{Of course we could also assume that the tensor term is proportional to $\delta_{(\rm K)}^{ij}$ and a scalar, in which case the tensor terms again reduce to the scalar terms.} and the vector terms reduce to the scalar terms.
These terms cover the usual case of a velocity perturbation that happens to align with the line of sight in projection.
If instead it is assumed that $\epsilon_{(\rm v)}^{i}(\vs) \propto \hat{z}^i$ and $\epsilon_{(\rm t)}^{ij}(\vs) \propto \hat{z}^i \hat{z}^j$ and a scalar field, we must go back to Eq.~\eqref{eq:add_stoch_0}, but the tensor and vector terms reduce to the scalar terms.
Such a stochastic field physically might correspond to a radiative galactic feedback mode with a line-of-sight component located between the observer and \Lya quasar backlight source (i.e., a change in ionization state uncorrelated with the local small-scale velocity field traced by an absorbing HI cloud).
The mixed tensor case, $\epsilon_{(\rm t)}^{ij}(\vs) \propto k^i \hat{z}^j$, is, like the $k^i k^j$ case, 3rd-order in derivatives, which we neglect. 
We then obtain
\begin{align} \label{eq:add_stoch_2}
    \epsilon^{(\rm{tot})}(\vs) &= \epsilon^{(\rm s,1)}(\vs) + k^2\epsilon^{(\rm s,2)}(\vs) + k_\parallel^2 \epsilon^{(\rm s,3)}(\vs),
\end{align}
where each $\epsilon^{(\rm s,i)}(\vs)$ is an effective scalar stochastic field.

This derivation, purely considering $SO(2)$ symmetry, does not require the use of the density-preserving real-to-redshift space mapping or the addition of stochastic fields for each of the contact operators appearing in the usual (perturbative) treatment \cite{Perko:2016}.
This allows us to rule out a subset of the terms in Eq.~\eqref{eq:add_stoch_0} 
based on assumed line-of-sight isotropy at the field level, while there the authors appealed to symmetry in real space coordinate correlators to argue that $\epsilon_{(\rm v)}^i \propto k^i$, $\epsilon_{(\rm t)}^{ij} \propto \delta_{(\rm K)}^{ij}$ and a scalar field\footnote{The use of $\delta_{(\rm K)}^{ij}$, and actually, \textit{any} tensor or vector stochasticity is totally redundant for obtaining the correct leading-derivative form of the (effective) stochasticity. However, the development of this expression from symmetry principles is clear and immediately generalizes to higher orders with further enumeration of powers of $k^i,\hat{z}^j$.}.
Both approaches, however, produce the expected power spectrum of the stochastic fields at this order.

Finally, we must consider the second-order stochastic contributions (neglecting higher derivatives for these terms)
\begin{align} \label{eq:add_stoch_3}
    \epsilon^{(\rm{tot})}(\vs) &= \epsilon^{(\rm s,1)}(\vs) + k^2\epsilon^{(\rm s,2)}(\vs) + k_\parallel^2 \epsilon^{(\rm s,3)}(\vs) \\
     &~ + \epsilon_{\delta}(\vs)\delta(\vs) + \epsilon_{\delta^\parallel}(\vs)\delta^\parallel(\vs),
\end{align}
where, as in Ref.~\cite{Desjacques2018}, we have introduced a stochastic contribution generated by $\delta^\parallel$ for the $SO(2)$-based bias expansion in addition the standard $\epsilon_\delta$ term.
While these terms are required for consistency at 2nd order, we note that, unlike the case of galaxies, the amplitude of the shot noise contributions (assuming Poisson statistics) are extremely small \cite{Ivanov2024,deBelsunce:2024rvv}\footnote{
In the full expression for the bias expansion, there are also displaced stochastic fields of the form $\psi^i\partial_iX$ where $X = \{\epsilon,\epsilon^\parallel\}$, that arise at quadratic order \cite{DJS_review,chen2020_lpt_rsd}:
\begin{align} \label{eq:add_stoch_4}
\psi^i \partial_i \epsilon, \psi^i \partial_i \epsilon^\parallel,\psi^i\epsilon^{\rm v}_i,...
\end{align}
These are often neglected and argued to be small since the tracer nonlocality scale is small for the usual case of galaxies and halos, but we list them here for completeness. }.

\subsection{\label{app_sub:restore_los} Recovering redshift-space galaxies}
The case of redshift-space galaxies is a special case of the general $SO(2)$ tracer we have worked with so far.
By applying the additional assumptions included in the usual galaxy real-to-redshift-space transformation, we can reduce the $SO(2)$ bias expansion (eqn.~\ref{eq:app_F_2d}) to this special case.
The quadratic bias parameters can be fixed to those of galaxies (or any 3D tracers) that have been mapped from real space into redshift space by imposing momentum conservation and mass conservation on the anisotropic parts of the $SO(2)$ kernel.
The usual isotropic operators in 3D $F_2$, $G_2$, $\mathcal{G}_2$ (and so the 2nd-order matter density/velocity fields) obey mass and momentum conservation  \cite{DAmico:2021_lss_bootstrap}, but the quadratic kernel in the bias expansion of galaxies in 3D does not satisfy number (mass) conservation as there is no physical expectation that number/mass is conserved for general tracers.
However, the real-to-redshift-space transformation is generated by the requirement of number conservation\footnote{This is the global number conservation enforced by eqn.~\eqref{eqn:rsd_density}, not the more restrictive local conservation $\rho(\vs)\mathrm{d}^3\vs = \rho(\vx)\mathrm{d}^3\vx$, which additionally requires a 1-to-1 mapping from real-to-redshift space that we do not assume.}.
Therefore, we can reduce the $SO(2)$ kernels to the case of galaxies by forcing the \textit{anisotropic} pieces of the more general $SO(2)$ kernel to satisfy mass and momentum conservation.
Specifically, this allows us to recover the $Z_2$ kernel from the $K_2$ kernel (See Appendix~\ref{app:consv_Z2} for details.).

The $Z_2$ kernel for real-to-redshift space mapped galaxies follows from explicitly applying (global) mass/number conservation 
and perturbatively expanding the exponential map applied to the line-of-sight matter velocity. 
This anisotropic part of this kernel also respects number conservation and the whole kernel satisfies momentum conservation. 
In Appendix~\ref{app:consv_Z2}, we show that by imposing mass/number and momentum conservation on the anisotropic part of the $K_2$ kernel for a $SO(2)$ tracer, we fix all of the bias parameters except for $b_1, b_\parallel$ and two of the quadratic bias parameters, $b_2$, $b_{\mathcal{G}_2}$, as expected for the case of redshift-space galaxies.

Some further discussion of the $b_\parallel \delta^\parallel$ term (with remaining free parameter $b_\parallel$) is warranted to understand how we recover the case of redshift-space galaxies from the $SO(2)$ expansion at linear order.
For a general $SO(2)$ tracer, $\delta^\parallel$ is one of operators allowed by the symmetries.
Its bias $b_\parallel$ is therefore unconstrained by the equivalence principle, as the operator has nothing to do with velocities or long-wavelength displacements.
For redshift-space galaxies, the \textit{only} source of LOS-dependence is $\vpsi_{\mathrm{RSD}}$ from the parallel components of galaxy peculiar velocities.
In the \textit{specific model} of RSD galaxies, we identify $b_\parallel \delta^\parallel = \eta_g = \partial_\parallel u_{g\parallel}$\footnote{With $u \equiv \frac{v}{\mathcal{H}}$.}, and, since velocities, and therefore RSD displacement, must obey the equivalence principle, the velocity must be unbiased $u_{g\parallel} = b_vu_{\parallel} = 1\cdot u_{\parallel}$, and we recover the Kaiser formula upon substituting the linear theory expression for $u$.

Of course, by \textit{ad hoc} equating the standard expression for redshift-space galaxies with Eq.~\eqref{eq:app_F_2d}, we also recover the expected values of the line-of-sight bias parameters $\frac{1}{\mathcal{H}f} b_{\parallel} = b_{\eta}=1,\frac{1}{\mathcal{H}f}b_{\delta\parallel} = b_{\delta\eta}=b_1,\frac{1}{(\mathcal{H}f)^2} b_{\parallel^2} = b_{\eta^{2}}=1$ as well as $b_{(KK)_{\parallel}}=b_{\Pi_{2}^{\parallel}}=0$.
Simply performing angular averages of over the line-of-sight vector also restores the case of \textit{real-space} galaxies, as the LOS-dependent operators vanish \cite{Desjacques2018}.

\section{Skew spectrum formalism} \label{sec:Lya_skewspectra}
To compute the \Lya skew spectra, we first write out the unsymmetric bispectrum (following Ref.~\cite{Schmittfull:2021}) at tree-level to account for all relevant contributions\footnote{We neglect stochastic contributions given that the Poisson expectation for the \LyaF is negligible (while this is a good approximation, see Ref.~\cite{deBelsunce:2025bqc} for a discussion of non-Poisson \LyaF stochasticity in the context of field level modeling). 
We also neglect higher-derivative terms under the assumption that the relevant scale for ``formation'' of \LyaF flux fluctuations is small, though this is not guaranteed given the possibility of UV background fluctuations \cite{Cabass2019:uvb_grad}. }.
These contributions are not the same as those for galaxies due to the presence of line-of-sight dependent terms. 
The presence of $b_\eta,b_{\delta\eta}$ adds coefficients to the terms in the 
galaxy bispectrum without line-of-sight dependent terms, while the $(KK)_\parallel,\Pi^{[2]}_\parallel$ terms contribute new $f^0$ terms to the fluctuation field. Grouping the terms by the growth factor $f$ as in Ref.~\cite{Schmittfull:2021}, we have:
\begin{equation}
  \label{eq:B_unsymm}
  B_\mathrm{FFF}^\mathrm{unsym}(\vk_1,\vk_2;\vk_3,\hat{\mathbf{z}}) =
2P_\mathrm{mm}(k_1)P_\mathrm{mm}(k_2)
 \sum_{n=0}^4 B^{f^n}(\vk_1,\vk_2;\hat{\mathbf{z}}),
\end{equation}
where $B^{f^n} \propto f^n$ and the fully symmetric bispectrum is given as the sum over the permutations 
\begin{equation}
    B_\mathrm{FFF}^\mathrm{sym}(\vk_1,\vk_2;\vk_3,\hat{\mathbf{z}}) = B_\mathrm{FFF}^\mathrm{unsym}(\vk_1,\vk_2;\vk_3,\hat{\mathbf{z}}) +
    B_\mathrm{FFF}^\mathrm{unsym}(\vk_1,\vk_3;\vk_2,\hat{\mathbf{z}}) + B_\mathrm{FFF}^\mathrm{unsym}(\vk_2,\vk_3;\vk_1,\hat{\mathbf{z}})\,,  
\end{equation}
with $\vk_3=-\vk_2-\vk_3$. This gives,
\begin{align}
\label{eq:f0}
  B^{f^0}(\vk_1, \vk_2;\hat{\mathbf{z}}) =&\;
 f^0\bigg\{ b_1^3
  F_2(\vk_1,\vk_2)
+ b_1^2\frac{b_2}{2} + b_1^2 b_{\mathcal{G}_2} \mathcal{G}_2(\vk_1,\vk_2) + b_1^2 \textcolor{blue}{b_{(KK)_\parallel}} (KK)_\parallel(\vk_1,\vk_2;\hat{\mathbf{z}}) + b_1^2\textcolor{blue}{b_{\Pi^{[2]}_\parallel}} \Pi^{[2]}_\parallel(\vk_1,\vk_2;\hat{\mathbf{z}}) \bigg \}, \\
\label{eq:f1}
  B^{f^1}(\vk_1, \vk_2;\hat{\mathbf{z}}) =&\;
f^1 \bigg\{\textcolor{red}{b_{\eta}} b_1^2 \left[ \left(\frac{k_{1\parallel}^{2}}{k_{1}^{2}} + \frac{k_{2\parallel}^{2}}{k_{2}^{2}}  \right)F_2(\vk_1,\vk_2) +\frac{k_{3\parallel}^{2}}{k_{3}^{2}}G_{2}(\vk_1,\vk_2)\right]
+\textcolor{red}{b_{\eta}}\frac{b_{1}b_2}{2}\left(\frac{k_{1\parallel}^{2}}{k_{1}^{2}} 
+ \frac{k_{2\parallel}^{2}}{k_{2}^{2}}  \right)\\ 
\nonumber
&\qquad +\textcolor{red}{b_{\eta}}b_{1} b_{\mathcal{G}_2}\left(\frac{k_{1\parallel}^{2}}{k_{1}^{2}} + \frac{k_{2\parallel}^{2}}{k_{2}^{2}}  \right) \mathcal{G}_2(\vk_1,\vk_2)  +\textcolor{red}{b_{\delta \eta}}\frac{b_{1}^{2}}{2}\left(\frac{k_{1\parallel}^{2}}{k_{1}^{2}} + \frac{k_{2\parallel}^{2}}{k_{2}^{2}}  \right) + \frac{b_{1}^{3}}{2} k_{1\parallel}k_{2\parallel}\left(\frac{1}{k_{1}^{2}} + \frac{1}{k_{2}^{2}}\right)\\
\nonumber
&\qquad +\textcolor{red}{b_{\eta}} b_{1} \left(\frac{k_{1\parallel}^{2}}{k_{1}^{2}} + \frac{k_{2\parallel}^{2}}{k_{2}^{2}} \right) \left( \textcolor{blue}{b_{(KK)_\parallel}} (KK)_\parallel(\vk_1,\vk_2;\hat{\mathbf{z}})
+ \textcolor{blue}{b_{\Pi^{[2]}_\parallel}} \Pi^{[2]}_\parallel(\vk_1,\vk_2;\hat{\mathbf{z}}) 
\right)
\bigg\},\nonumber \\
\label{eq:f2}
  B^{f^2}(\vk_1, \vk_2;\hat{\mathbf{z}}) =&\;
f^2 \bigg\{ \textcolor{red}{b_{\eta}}^{2} b_1 \left(\frac{k_{1\parallel}^{2}k_{2\parallel}^{2}}{k_{1}^{2}k_{2}^{2}}  F_2(\vk_1,\vk_2) +  \left(\frac{k_{1\parallel}^{2}}{k_{1}^{2}} + \frac{k_{2\parallel}^{2}}{k_{2}^{2}} \right)\frac{k_{3\parallel}^{2}}{k_{3}^{2}}G_{2}(\vk_1,\vk_2) \right)\\
&\qquad + \textcolor{red}{b_{\eta}}^{2} \frac{b_2}{2} \frac{k_{1\parallel}^{2}k_{2\parallel}^{2}}{k_{1}^{2}k_{2}^{2}} + \textcolor{red}{b_{\eta}}^{2} 
 b_{\mathcal{G}_2}\frac{k_{1\parallel}^{2}k_{2\parallel}^{2}}{k_{1}^{2}k_{2}^{2}}\mathcal{G}_2(\vk_1,\vk_2)  \nonumber \\
\nonumber
&\qquad + \textcolor{red}{b_{\eta}b_{\delta \eta}} \frac{b_1}{2} \left(\frac{k_{1\parallel}^{2}}{k_{1}^{2}} + \frac{k_{2\parallel}^{2}}{k_{2}^{2}} \right)^{2} + \textcolor{red}{b_{\eta}}\frac{b_{1}^{2}}{2} \left(\frac{k_{1\parallel}^{2}}{k_{1}^{4}} + \frac{k_{2\parallel}^{2}}{k_{2}^{4}} + 2 \frac{k_{1\parallel}^{2}+k_{2\parallel}^{2}}{k_{1}^{2} k_{2}^{2}}\right) k_{1\parallel}k_{2\parallel} + b_{1}^{2} \textcolor{red}{b_{\eta^2}}\frac{k_{1\parallel}^{2}k_{2\parallel}^{2}}{k_{1}^{2} k_{2}^{2}}\\
\nonumber
&\qquad + \textcolor{red}{b_{\eta}}^{2} \frac{k_{1\parallel}^{2}k_{2\parallel}^{2}}{k_{1}^{2}k_{2}^{2}} \left( \textcolor{blue}{b_{(KK)_\parallel}} (KK)_\parallel(\vk_1,\vk_2;\hat{\mathbf{z}})
+ \textcolor{blue}{b_{\Pi^{[2]}_\parallel}} \Pi^{[2]}_\parallel(\vk_1,\vk_2;\hat{\mathbf{z}}) 
\right)
\bigg\},\nonumber\\
\label{eq:f3}
  B^{f^3}(\vk_1, \vk_2;\hat{\mathbf{z}}) =&\;
f^3 \bigg\{ \textcolor{red}{b_{\eta}} \textcolor{red}{b_{\eta^2}} b_1 \left(\frac{k_{1\parallel}^{2}}{k_{1}^{2}} + \frac{k_{2\parallel}^{2}}{k_{2}^{2}} \right)  \frac{k_{1\parallel}^{2}k_{2\parallel}^{2}}{k_{1}^{2}k_{2}^{2}}  + \frac{1}{2} \textcolor{red}{b_{\eta}}^2  b_1 \frac{1}{k_1^4 k_2^4}\left[ k_{1\parallel}^{5}k_{2\parallel} k_{2}^{2} + k_{1\parallel}k_{2\parallel}^5 k_{1}^{2} + 2 k_{1\parallel}^{3}k_{2\parallel}^{3}(k_{1}^{2}+k_{2}^{2}) \right]\\ 
\nonumber
&\qquad + \textcolor{red}{b_{\eta}}^{2} \frac{\textcolor{red}{b_{\delta\eta}}}{2}\frac{k_{1\parallel}^{2}k_{2\parallel}^{2}}{k_{1}^{2}k_{2}^{2}}\left(\frac{k_{1\parallel}^{2}}{k_{1}^{2}} + \frac{k_{2\parallel}^{2}}{k_{2}^{2}} \right)
+\textcolor{red}{b_{\eta}}^{3}  \frac{k_{1\parallel}^{2} k_{2\parallel}^{2}}{k_{1}^{2} k_{2}^{2}} \frac{k_{3\parallel}^{2}}{k_{3}^{2}}G_{2}(\vk_1,\vk_2)
\bigg\},
\\
\label{eq:f4}
  B^{f^4}(\vk_1, \vk_2;\hat{\mathbf{z}}) =&\;
f^4 \bigg\{
  \textcolor{red}{b_{\eta}}^2\textcolor{red}{b_{\eta^2}} \frac{k_{1\parallel}^{4}k_{2\parallel}^{4}}{k_{1}^{4} k_{2}^{4}} +\textcolor{red}{b_{\eta}}^3\frac{1}{2} \frac{k_{1\parallel}^{2}k_{2\parallel}^{2}}{k_{1}^{2} k_{2}^{2}} \left( \frac{k_{1\parallel}k_{2\parallel}^{3}}{k_{1}^{2} k_{2}^{2}} + \frac{k_{1\parallel}^{3}k_{2\parallel}}{k_{1}^{2} k_{2}^{2}} \right)
\bigg\} \,, 
\end{align}
where terms that are completely new compared to Ref.~\cite{Schmittfull:2021} are highlighted in \textcolor{blue}{blue} and terms now accompanied by a bias parameter are highlighted in \textcolor{red}{red}.
Note that the new terms enter without any factors of the growth rate $f$, and so contribute similarly to the usual SPT second-order terms. 
As expected \cite{Desjacques2018}, the \Lya bispectrum reduces to the galaxy bispectrum in the absence of line-of-sight dependent terms 
with $\textcolor{red}{b_{\eta}}=1, \textcolor{red}{b_{\eta^{2}}}=1,\textcolor{red}{b_{\delta\eta}}=b_{1}$ and $\textcolor{blue}{b_{(KK)_\parallel}} = \textcolor{blue}{b_{\Pi^{[2]}_\parallel}} = 0$

\subsection{\Lya Skew-spectrum operators} \label{sec:skewspec_operators}
Using the expressions in Eqs.~\eqref{eq:f0}-\eqref{eq:f4}, we can now enumerate the new skew spectra operators\footnote{Arguments are suppressed, but a general skew spectrum $\mathcal{S}_n$ has arguments of the form $\mathcal{S}_{n}(X(\vk_1),Y(\vk_2),\vk_1,\vk_2;\hat{\mathbf{z}})$ for fields $X, Y$. Configuration space quadratic operators take the same symmetrized form of Ref.~\cite{Schmittfull:2021}, e.g. $F_{2}[X,Y](\vx) = \int_{\vk} e^{i\vk \cdot \vx} \frac12 \int_{\vq} \left(X(\vq)Y(\vk-\vq) + Y(\vq)X(\vk-\vq)\right) F_{2}(\vk,\vk-\vq)$ }, of which there are 26:
\begin{align} \label{eq:S0}
    b_{1}^{3}f^{0}&: \, \mathcal{S}_{1} = F_2\left[\td,\td\right]\\
    b_{1}^{2}b_{2}f^{0}&: \, \mathcal{S}_{2} = \td^2\\
    b_{1}^{2}b_{\mathcal{G}_2}f^{0}&: \, \mathcal{S}_{3} = \mathcal{G}_2\left[\td,\td\right]\\
    b_{1}^{2}\textcolor{blue}{b_{(KK)_\parallel}}f^{0}&: \, \mathcal{S}_{4} = (KK)_{\parallel}[\delta,\delta]\\
    b_{1}^{2}\textcolor{blue}{b_{\Pi^{[2]}_\parallel}}f^{0}&: \, \mathcal{S}_{5} = \Pi^{[2]}_{\parallel}[\delta,\delta]\\
    b_{1}^{3}f^{1}&: \, \mathcal{S}_{6} = \hat{z}^{i}\hat{z}^{j} \partial_i \delta \frac{\partial_j}{\nabla^2}\delta \\
    b_{1}^{2}\textcolor{red}{b_{\eta}}f^{1}&: \, \mathcal{S}_{7} = G_{2}^{\parallel}[\delta,\delta] + 2 F_{2}[\delta^{\parallel},\delta]\\
    b_{1}b_{2}\textcolor{red}{b_{\eta}}f^{1}&: \, \mathcal{S}_{8} = \delta^{\parallel} \delta\\
    b_{1}b_{\mathcal{G}_2}\textcolor{red}{b_{\eta}}f^{1}&: \, \mathcal{S}_{9} = \mathcal{G}_{2}[\delta^{\parallel},\delta]\\
    b_{1}\textcolor{red}{b_{\eta}}\textcolor{blue}{b_{(KK)_\parallel}}f^{1}&: \, \mathcal{S}_{10} = (KK)_{\parallel}[\delta^{\parallel},\delta] \\
    b_{1}\textcolor{red}{b_{\eta}}\textcolor{blue}{b_{\Pi^{[2]}_\parallel}}f^{1}&: \, \mathcal{S}_{11} = \Pi_{\parallel}^{[2]}[\delta^{\parallel},\delta]\\
    b_{1}^{2}\textcolor{red}{b_{\delta\eta}}f^{1}&: \, \mathcal{S}_{12} = \delta^{\parallel} \delta \\
    b_{1}(\textcolor{red}{b_{\eta}})^2 f^{2}&: \, \mathcal{S}_{13} = 2G_{2}^{\parallel}[\delta^{\parallel},\delta] + F_{2}[\delta^{\parallel},\delta^{\parallel}]\\
    b_{2}(\textcolor{red}{b_{\eta}})^2 f^{2}&: \, \mathcal{S}_{14} = \left(\delta^{\parallel} \right)^{2}\\
    b_{\mathcal{G}_2}(\textcolor{red}{b_{\eta}})^2 f^{2}&: \, \mathcal{S}_{15} = \mathcal{G}_2[\delta^{\parallel},\delta^{\parallel}]\\
    (\textcolor{red}{b_{\eta}})^2 \textcolor{blue}{b_{(KK)_\parallel}} f^{2}&: \, \mathcal{S}_{16} = (KK)_{\parallel}[\delta^{\parallel},\delta^{\parallel}]\\
    (\textcolor{red}{b_{\eta}})^2 \textcolor{blue}{b_{\Pi^{[2]}_\parallel}} f^{2}&: \, \mathcal{S}_{17} = \Pi_{\parallel}^{[2]}[\delta^{\parallel},\delta^{\parallel}]\\
    b_{1}\textcolor{red}{b_{\eta}}\textcolor{red}{b_{\delta \eta}}f^{2}&: \, \mathcal{S}_{18} = \delta^{\parallel\parallel}\delta + 2\left(\delta^{\parallel}\right)^2\\
    b_{1}^{2}\textcolor{red}{b_{\eta}}f^{2}&: \, \mathcal{S}_{19} = \hat{z}^{i} \hat{z}^{j} \left[\left(\partial_{i}\delta\right)\left(\frac{\partial_{j}}{\nabla^2}\delta^{\parallel}\right) + 2\left(\partial_{i}\delta^{\parallel}\right)\left(\frac{\partial_{j}}{\nabla^2}\delta\right) \right]\\
    b_{1}^{2}\textcolor{red}{b_{\eta^2}} f^{2}&: \, \mathcal{S}_{20} = \left(\delta^{\parallel} \right)^2\\
    b_1 (\textcolor{red}{b_{\eta}})^2f^{3}&: \, \mathcal{S}_{21} = \hat{z}^{i} \hat{z}^{j} \left[ \left(\partial_{i}\delta^{\parallel\parallel}\right)\left(\frac{\partial_{j}}{\nabla^2}\delta\right) + 2 \left(\partial_{i}\delta^{\parallel}\right)\left(\frac{\partial_{j}}{\nabla^2}\delta^{\parallel}\right)\right]\\
    b_1 \textcolor{red}{b_{\eta} b_{\eta^{2}}}f^{3}&: \, \mathcal{S}_{22} = \delta^{\parallel \parallel} \delta^{\parallel}\\
    (\textcolor{red}{b_{\eta}})^2 \textcolor{red}{b_{\delta\eta}}f^{3}&: \, \mathcal{S}_{23} = \delta^{\parallel \parallel} \delta^{\parallel}\\
    (\textcolor{red}{b_{\eta}})^3f^{3}&: \, \mathcal{S}_{24} = G_{2}^{\parallel}[\delta^{\parallel},\delta^{\parallel}]\\
    (\textcolor{red}{b_{\eta}})^3f^{4}&: \, \mathcal{S}_{25} = \hat{z}^{i} \hat{z}^{j}\left(\partial_{i}\delta^{\parallel\parallel}\right)\left(\frac{\partial_j}{\nabla^2}\delta^{\parallel}\right)\\
    (\textcolor{red}{b_{\eta}})^2 \textcolor{red}{b_{\eta^2}}f^{4}&: \, \mathcal{S}_{26} = \left(\delta^{\parallel \parallel}\right)^2. \label{eq:S26}
\end{align}

These reduce to the 14 skew spectra terms for galaxies of Ref.~\cite{Schmittfull:2021} when $\textcolor{red}{b_{\eta}}=1, \textcolor{red}{b_{\eta^{2}}}=1,\textcolor{red}{b_{\delta\eta}}=b_{1}$ and $\textcolor{blue}{b_{(KK)_\parallel}} = \textcolor{blue}{b_{\Pi^{[2]}_\parallel}} = 0$, which are also used in Refs.~\cite{2023JCAP...03..045H,2024JCAP...05..011C}.\footnote{Ref.~\cite{Dizgah_2020} assumes measuring the 14 skew spectra for galaxies extracts the same information on galaxy bias parameters and growth rate as the full bispectrum. Here, we assume this to hold for the \Lya forest as well.} The corresponding skew spectra in Fourier space are given in Appendix~\ref{app:kspace_skew}.

\subsection{Computation of anisotropic skew spectra} \label{sec:computation_ss}
\begin{figure}
    \centering
    \includegraphics[width=\linewidth]{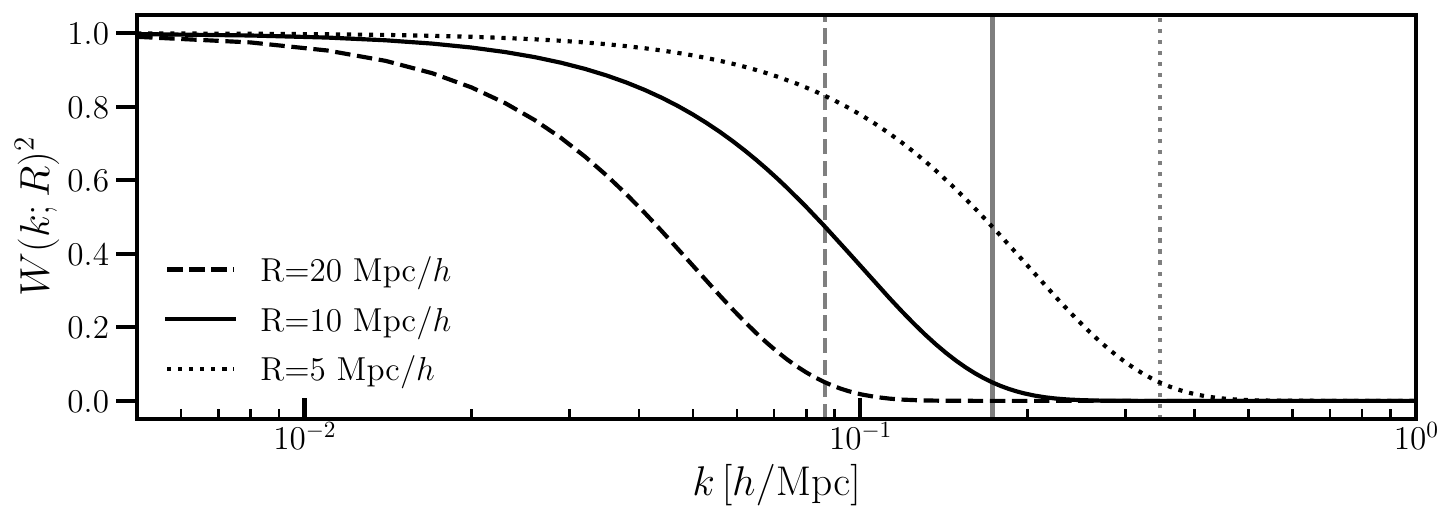}
    \caption{Squared isotropic smoothing kernel $W(k; R)=\exp{-\left(\frac12 k^2R^2\right)}$ for wavenumber $k$ and smoothing radii $R=20\hinvMpc$ shown as a dashed,  $R=10\hinvMpc$ as a solid and  $R=5\hinvMpc$ as a dotted black line. The vertical gray lines indicate the scales where power is suppressed by a factor of 20. 
    }
    \label{fig:smoothing_kernel}
\end{figure}

We compute the theoretical anisotropic skew spectra as follows:
\begin{equation} \label{eq:skewspectrum}
    P_{\mathcal{S}^{\ell}_n}(k) = \frac{2\ell+1}{2}\int\frac{\diff\hat{\kvec}}{4\pi} \mathcal{L}_{\ell}(\mu_k)\int_{\vq} \mathcal{S}_{n}(\vq,\vk-\vq) B(\vq,\vk-\vq,-\vk),
\end{equation}
with quadratic real-space kernels $\mathcal{S}_n$ given in Eqs.~\eqref{eq:S0}-\eqref{eq:S26} (and the Fourier space kernels in Appendix~\ref{app:kspace_skew}), and Legendre multipoles $\mathcal{L}_{\ell}(\mu) = \{1,\, \frac12 (3\mu^2-1)\}$, for the monopole ($\ell=0$) and quadrupole ($\ell=2$), respectively. In the present work, we truncate the multipole expansion at $\ell=2$. We choose the coordinate system \cite{Schmittfull:2021}:
\begin{equation}
    \hat{\mathbf{z}} = (0,0,1),\qquad \hat{\mathbf{k}}=\left(0, \sqrt{1-\mu_k^2},\mu_k\right),\qquad \hat{\mathbf{q}}=\left(\sqrt{1-\mu_q^2}\cos\phi_q,\sqrt{1-\mu_q^2}\sin\phi_q,\mu_q\right),
\end{equation}
where $\mu_k=\hat{\vk}\cdot\hat{\mathbf{z}}$, $\mu_q=\hat{\vq}\cdot\hat{\mathbf{z}}$, $\cos(\phi_q)=\hat{\vq}\cdot\hat{\mat{x}}$, and $\hat{\mat{z}}$ is the line-of-sight direction.\footnote{We use the publicly available \texttt{pycuba} package (\url{https://github.com/JohannesBuchner/PyMultiNest/}) to numerically evaluate the four-dimensional integral.} 
For performing consistency tests similar to those of Ref.~\cite{Schmittfull:2021}, we smooth the density field using a Gaussian kernel $\td_R(\vk) = \td(\vk) \exp{\left(-\frac12 k^2R^2\right)}$, illustrated in Fig.~\ref{fig:smoothing_kernel}. This effectively removes small-scale modes in Eq.~\eqref{eq:skewspectrum} above $k \approx 2/R\, \hMpcinv$ (a smoothing scale of $R=5,\, 10,\, 20 \hinvMpc$ corresponds to a suppression of power by a factor of 20 at $k=0.34,\, 0.17,\, 0.09\, \hMpcinv$). Our baseline results use $R=10 \hinvMpc$.\footnote{Note that only the fields entering the quadratic operator $\mathcal{S}_n$ are smoothed, while the field entering linearly is not smoothed. The smoothing scale is well below the non-linear scale, $\knl$, for the \Lya forest. It is defined where the dimensionless power spectrum reaches unity, i.e. $\Delta^2 = k^3P_{\rm lin}(k,z)/(2\pi^2)\sim 1$, corresponding to  $\knl = 4.1 \hMpcinv$ at the redshift and cosmology of the \abacus simulation. Beyond which an EFT model should be treated as phenomenological.}
(A simple isotropic smoothing of the 3D flux 
field is of course not possible for real 
\LyaF\ observations -- we address this below
by using a purely radial smoothing kernel.)

To measure the skew spectra, we follow the approach of Ref.~\cite{Schmittfull:2021} and implement\footnote{We use the publicly available codes \textsc{Nbodykit} (\url{https://nbodykit.rtfd.io}) and \textsc{skewspec} (\url{https://github.com/mschmittfull/skewspec}).} the Fourier-space skew spectra given in appendix~\ref{app:kspace_skew}. We apply each filter to the Fourier transformed input flux field, $\td_F$. We Fourier transform it back and multiply a second copy of the filtered input field. The resulting quadratic (and smoothed) field is then correlated with the original input field of which we measure the power spectra
\begin{equation}
    \label{eq:measured_skew}
    \hat{P}_{\mathcal{S}_n^\ell}(k) = \frac{1}{V} \int \frac{d\hat{\vk}}{4\pi}\mathcal{L}_\ell(\mu_k) \delta_{R}(-\vk) \int_{\vq}\mathcal{S}_{n}(\vq,\vk-\vq) \delta_{R}(\vq)\delta_{R}(\vk-\vq).
\end{equation}

\section{Results on simulations} \label{sec:simulations}
In this section, we present consistency tests of the anisotropic skew spectrum methodology, described in Sec.~\ref{sec:Lya_skewspectra}, on two sets of simulations: (i) synthetic three-dimensional \Lya fields using perturbation theory up to second order (denoted by 2SPT); and (ii) on large \Lya forest mocks from the \abacus suite of mocks \cite{Hadzhiyska:2023}. For the 2SPT mocks, we generate a Gaussian random field, $\td(\vk)$, with a linear matter power spectrum, $P_{\rm lin}(k)$, using a fiducial \textit{Planck} 2018 cosmology \cite{Aghanim:2018eyx}: $\Omega_b h^2 = 0.02237$, $\Omega_c h^2 = 0.12$, $h = 0.6736$, $A_s = 2.0830 \times 10^{-9}$, $n_s = 0.9649$, $w_0 = -1$, $w_a = 0$ at redshift $z=2.5$ as input. Following \cite{Schmittfull:2021}, we use Eq.~\eqref{eq:F_full} to generate $\td_F$. The bias parameters for this test are set to the \textit{measured} bias parameters of the \Lya forest from \abacus simulations, tabulated in the third row of Table~\ref{tab:abacus_models}. These are obtained by fitting the one-loop \Lya forest EFT power spectrum to the simulations \cite{Abacus_BAO_Lya:2025}. Evaluating the skew spectra on field-level perturbative mocks and comparing to perturbation theory predictions provides a consistency check of the implementation.\footnote{Note that when setting the bias parameters to $\{b_1,\, b_\eta,\, b_2,\, b_{\mathcal{G}_2},\, b_{\delta \eta},\, b_{\eta^2},\, b_{(KK)_\parallel},\, b_{\Pi^{[2]}_\parallel}\} = \{1,\, 1,\,0,\,0,\,1,\,1,\,0,\,0\}$ we recover the dark matter skew spectra redshift space.} We do not expect perfect agreement between the theoretical prediction at tree-level and the measured numerical skew spectra from the 2SPT field, as the so-computed numerical skew spectra will include terms that are higher order in perturbation theory (e.g. $\langle \delta_2 ~\delta_2 ~\delta_2\rangle$, $\langle \delta_2~ \delta \eta ~\eta_2\rangle$,...). However, on large scales, we expect that the numerical skew spectra computed from the 2SPT field and the tree-level theory should match. This can be regulated through the smoothing scale $R$.  In fact, due to the lower bias of the \Lya forest and lower non-linearity at higher redshift we expect to reach a higher $k_{\rm max}$ compared to galaxies. 

Whilst the 2SPT fields validate our theory pipeline and methodology, a more stringent test of our framework is to apply it to more realistic mocks. Therefore, we use the \abacus simulations which have the \Lya forest painted on top of them \cite{Hadzhiyska:2023,Abacus_BAO_Lya:2025}. This set of simulations is based on a fiducial \textit{Planck} 2018 cosmology and contains 6912$^3$ particles with a box length of $2\hinvGpc$, each with a mass of $M_{\rm part} = 2.1 \times 10^9$. Note that each model in Table~\ref{tab:abacus_models} corresponds to a different implementation of the fluctuating Gunn-Peterson approximation (FGPA; we refer the reader to Ref.~\cite{Hadzhiyska:2023} for more details and see Ref.~\cite{Abacus_BAO_Lya:2025} for a discussion in the context of the one-loop EFT \Lya forest power spectrum). The snapshot is at redshift $z = 2.5$ and we have six simulations for each model each with two lines of sights ($y$ and $z$) through the boxes. 
The 2SPT fields have been generated at a resolution of $N_{\rm c}=512$ cells per box. Whilst the \abacus simulations with the transmitted flux spectra have been constructed on a grid with $N_{\rm c}=6912$ cells\footnote{We have verified that our results are robust to the chosen resolution.} , we 
then degrade them to the same resolution of $N_{\rm c}=512$ which results in an implicit smoothing of $\approx 4 \hinvMpc$.

In the following, we compare two skew spectrum analyses with increasing levels of realism: in Sec.~\ref{sec:results_I} we use the isotropic, Gaussian smoothing scheme introduced in Sec.~\ref{sec:computation_ss} 
and in Sec.~\ref{sec:results_III} we introduce shifted skew spectra with a purely radial smoothing avoiding locally applying quadratic operators to the field which, in turn, removes the need for renormalization. In Sec.~\ref{sec:p3d_delta2} we compute the skew spectrum $\mathcal{S}_2$ explicitly including a survey geometry using a weighted pair count estimator and forward modeling the window. The third approach is \textit{directly applicable} to observational data and the key result of the present work. 

We emphasize that we do not fit the skew spectra to the \abacus simulation outputs. Instead, we compute the theoretical predictions using best-fit parameters obtained from one-loop EFT fits to the \Lya power spectrum (see \cite{Abacus_BAO_Lya:2025} for details). Since cubic terms such as $b_\eta^3$ appear in the tree-level bispectrum but not in the power spectrum, the skew spectra are sensitive to different combinations of bias parameters. Consequently, perfect agreement between the theory curves and the measurements from the \abacus simulations is not expected. A dedicated fit of the skew spectra is deferred to future work.

\subsection{Idealized Scenario: isotropic Gaussian smoothing} \label{sec:results_I}
\begin{figure}
    \centering
    \includegraphics[width=1\linewidth]{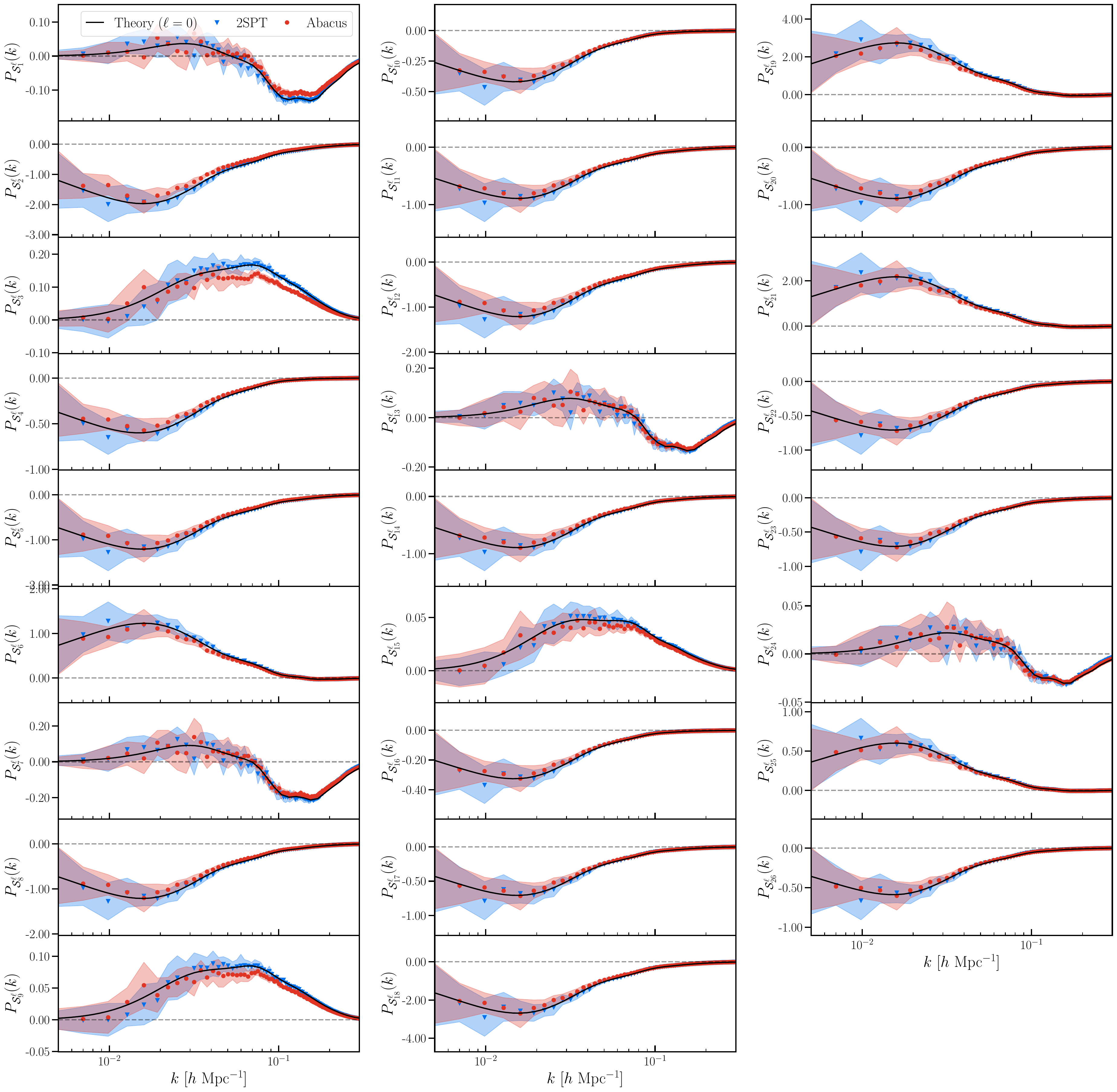}
    \caption{
    Consistency test of the skew spectrum monopole methodology and pipeline, comparing the 26 theory \Lya forest skew spectra (black solid line) to averages over 12 realizations of the synthetic three-dimensional \Lya fields using perturbation theory up to second order (2SPT; blue triangles)  and average over 12 realizations of the \abacus \Lya forest mock (red dots) for model III. The skew spectra are presented in units of volume, specifically $P_{\mathcal{S}^{\ell}_{i}}(k)$ in $[h^3 \, \mathrm{Mpc}^{-3}]$ for index $i$. For the theory and 2SPT fields we use the bias parameters given in Table~\ref{tab:abacus_models} which are obtained by fitting the one-loop EFT \Lya power spectrum to the \abacus simulations. We apply a Gaussian smoothing kernel to fields entering the quadratic operators with $R=10\hinvMpc$, corresponding to truncation of skew spectra (or bispectrum information) above $k \simgt  0.17 \hMpcinv$. The error bars are the root mean square between the realizations. Following baseline expectation, we find excellent agreement between the theory predictions and the 2SPT fields and agreement at the $\sim 2\sigma$ level with the \abacus data points given the different sensitivity to bias parameters. 
   }
    \label{fig:comparison_2SPT_abacus_model3_ell0}
\end{figure}

\begin{figure}
    \centering
    \includegraphics[width=1\linewidth]{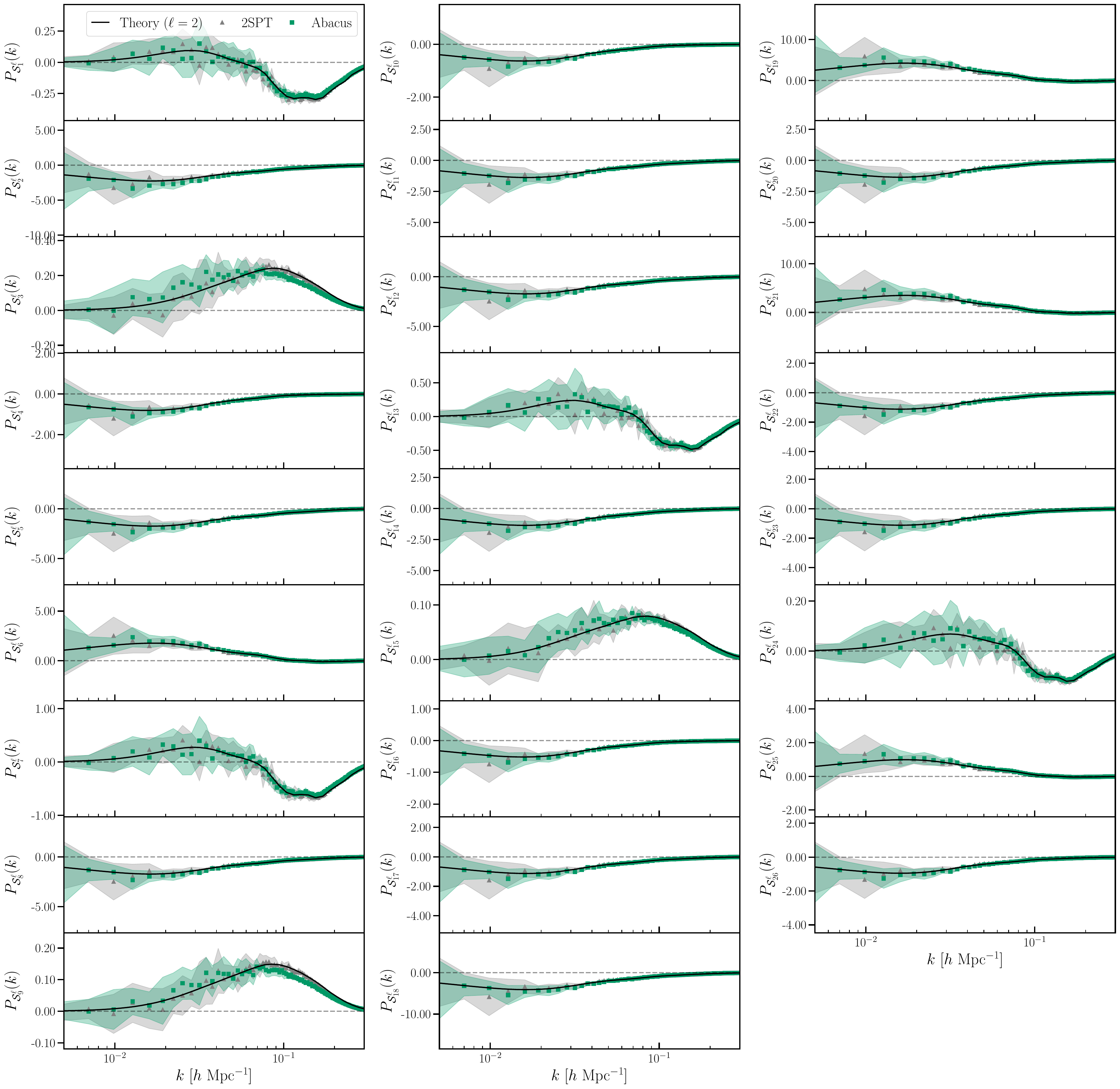}
    \caption{Same as Fig.~\ref{fig:comparison_2SPT_abacus_model3_ell0} for the quadrupole: comparison of 26 theory \Lya forest skew spectra (black solid line) to averages over 12 realizations of the synthetic three-dimensional \Lya fields using perturbation theory up to second order (2SPT; gray triangles)  and average over 12 realizations of the \abacus \Lya forest mock (green squares) for model III.  The error bars are the root mean square between the realizations.}
    \label{fig:comparison_2SPT_abacus_model3_ell2}
\end{figure}

In Figs.~\ref{fig:comparison_2SPT_abacus_model3_ell0} and \ref{fig:comparison_2SPT_abacus_model3_ell2} we compare the theoretically predicted anisotropic skew spectra using tree-level perturbation theory (black lines) to the measured ones from (i) synthetic three-dimensional \Lya fields using perturbation theory up to second order (blue lines) and (ii) from \abacus simulations. 
We find agreement at the $1$-$2\sigma$ level between the theory prediction and the measured skew spectra. The error bars are obtained from the square root of the variance between the $N=12$ simulations (two lines-of-sight and six realizations each). The Gaussian smoothing suppresses small-scale information beyond $k \simgt 0.17 \hMpcinv$. Whilst the theory prediction and the 2SPT measurements are in excellent agreement, we find differences at $1$-$2\sigma$ level at scales $k \sim 0.04-0.1 \hMpcinv$ to the \abacus measured skew spectra in $\mathcal{S}_{1,9,15}$. We emphasize that the plotted theory curves are generated using best-fit values for the bias parameters obtained from fitting the one-loop EFT power spectrum to \abacus simulations \cite{Abacus_BAO_Lya:2025} -- given the agreement between theory and 2SPT fields the discrepancies are mainly coming from different sensitivity to bias parameters of the one-loop power spectrum compared to skew spectra. Note that since for the \LyaF the RSD parameter is greater than unity, the quadrupole is larger than the monopole (in contrast to the case for galaxies). We choose $R=10\hinvMpc$ as our baseline smoothing scale, focusing on large (\textit{i.e.}~quasi-linear) scales where we expect the tree-level EFT model to work best. In App.~\ref{app:corr_mat} we investigate the degree of correlation between skew spectra in the covariance matrix.

Whilst the isotropic Gaussian smoothing presented here
is not applicable to real data, it validates the skew spectrum measurement and theory pipeline. As a first, na\"ive step towards applicability to real data one could apply a purely radial smoothing, \textit{i.e.}~using the same Gaussian kernel $\td_R(\vk) = \td(\vk) \exp{(-\frac12 k^2_{\parallel}R^2)}$. This, however, has the caveat that we are still squaring the field locally. 

When computing skew spectra, it is problematic to compute local products of the density field (e.g. $\delta^2(\vx)$), as this involves high-$k$ modes that are not under perturbative control (i.e., it requires renormalization of $\delta^2$; see, e.g.,~\cite{McDonald:2009}).
If one smooths the field isotropically,
and as performed, e.g., in \cite{2015PhRvD..91d3530S,Schmittfull:2021} then the impact of these modes is suppressed.
For the case of the \Lya forest flux fluctuations, the transverse sampling of the field is sparse for realistic datasets. In contrast to galaxies, for \Lya data it is not possible to smooth away the high-$k$ modes that have been aliased into the data, as one cannot determine which modes in the measurement come from high-$k$.  
In the line-of-sight direction, the data densely samples the underlying density field, so we are able to smooth the data to keep the line-of-sight dependent skew spectrum modes under control using $\td_{R_{\parallel}}(\vk) = \td(\vk) \exp{(-\frac12 k^2_{\parallel}R_\parallel^2)}$.
We present results using this anisotropic smoothing along with an enhanced skew spectrum statistic in the following Section.

\begin{table}[t!]
    \centering
    \setlength{\tabcolsep}{10pt}
    \begin{tabular}{ccccccccc}
        \hline
        \hline
        \vspace{-2ex} \\ 
        Model & $b_1$ & $b_\eta$ & $b_2$ & $b_{\mathcal{G}_2}$ & $b_{\delta \eta}$ & $b_{\eta^2}$ & $b_{(KK)_\parallel}$ & $b_{\Pi^{[2]}_\parallel}$ \\ [2ex]
        \hline \vspace{-2ex} \\ 
        I& -0.1485 & -0.1423 & -0.1231 & -0.0847 & 0.1088 & -0.3771 & -0.0751 & -0.3046 \\[1ex]
        II & -0.1314 & -0.1295 & -0.3316 & -0.2765 & 0.2631 & -0.3511 & \phantom{-}0.4841 & -0.2943 \\[1ex]
        III & -0.1337 & -0.2705 & -0.0353 & -0.0383 & 0.0811 & -0.0633 & -0.0164 & -0.2429 \\[1ex]
        IV& -0.1295 & -0.3041 & -0.0400 & -0.0320 & 0.0385 & -0.0167 & \phantom{-}0.0327 & -0.2578 \\[1ex]
        \hline
    \end{tabular}
    \caption{Mean best-fit values for the one-loop EFT parameters obtained from the \textsc{AbacusSummit} for each of the models one to four using a single simulation and line-of-sight. See Ref.~\cite{Abacus_BAO_Lya:2025} for a detailed discussion and presentation of the fitting methodology and Ref.~\cite{Hadzhiyska:2023} on specifics how \Lya forests are planted onto the $N$-body simulation.
    Note that these values use our sign convention rather than that of Ref.~\cite{Abacus_BAO_Lya:2025}. We choose Model III for this work as the linear bias parameters ($b_1$ and $b_\eta$) are closest to measurements from hydrodynamic simulations \cite{Chabanier:2024knr} and DESI data \cite{DESI_lya_2024}.}   
    \label{tab:abacus_models}
\end{table}

\subsection{Shifted skew spectra: non-local quadratic operations}\label{sec:results_III}

The skew spectra involve only quadratic operations that are computed at a single position.
This effectively corresponds to the squeezed limit of the \Lya bispectrum.
We can further enhance the information we can access from the bispectrum by probing other shapes -- to do this we introduce the \textit{shifted skew spectra}.
These statistics heuristically correspond to configuration-space triangles with two points located with some separation along one \Lya skewer and one point on a separate skewer

To motivate this, consider the three-point function 
\begin{align}
    \label{eqn:shift_skew_spec}
\langle \delta_F(\vs')\delta_F(\vs-\Delta\vs)\delta_F(\vs+\Delta\vs)\rangle,
\end{align}
which corresponds heuristically to two \Lya skewers separated by distance $|\vs'-\vs|$ where on one of the skewers we consider fluctuations separated along the line of sight $\hat{z}$ by $|\Delta\vs|$ - i.e. $\Delta\vs = \alpha \hat{z}$.
In Fourier space, this gives the bispectrum
\begin{align}
        \label{eqn:shift_skew_spec_k}
\langle \delta_F(\vk_1)\delta_F^{-\alpha}(\vk_2)\delta_F^{\alpha}(\vk_3)\rangle &= e^{-i\alpha k_{3\parallel}} e^{i\alpha k_{2\parallel}}  \langle \delta_F(\vk_1)\delta_F(\vk_2)\delta_F(\vk_3)\rangle\\
&=(2\pi)^3  e^{-i\alpha k_{3\parallel}}e^{i\alpha k_{2\parallel}} B_{FFF}(\vk_1,\vk_2,\vk_3)\delta^{(D)}(\vk_1+\vk_2+\vk_3)\nonumber\\
&=(2\pi)^3  e^{-i\alpha (k_{3\parallel}-k_{2\parallel})} B_{FFF}(\vk_1,\vk_2,\vk_3)\delta^{(D)}(\vk_1+\vk_2+\vk_3)\nonumber
\end{align}
as the Fourier flux overdensity field when shifted by a constant $\Delta \vs$ is
\begin{align}
    \label{eqn:shift_field}
    \delta_F^{\alpha}(\vk) &\equiv \int_\vs  e^{-i \vs \cdot \vs } \delta_F(\vs+\Delta \vs)
    = e^{-i k_\parallel \alpha} \delta_F(\vk)
\end{align}
where we introduced a scalar free parameter $\alpha$ that fixes the shift along the line of sight $\Delta \vs(\alpha,\hat{z}) = \alpha\hat{z}$.
We can then move this phase factor into each of the kernels $S_{n}$ corresponding to each skew spectrum, such that any quadratic field now is produced via a kernel
\begin{equation}
    \label{eqn:shifted_kernel}
    \mathcal{S}_{n,\alpha}(\vk_1,\vk_2,\hat{z}) =  \cos \left(\alpha (k_{1\parallel}- k_{2\parallel})\right)\mathcal{S}_{n}(\vk_1,\vk_2,\hat{z}),
\end{equation}
where we have symmetrized the shift and used Euler's identity. 
The parameter $\alpha$ allows us to generalize the usual skew spectra that involve local quadratic operations (\textit{i.e.}, $\alpha=0$) to a family of skew spectra where each value of $\alpha$ probes a different non-squeezed piece of the full bispectrum.
While this still falls short of the full set of triangles accessible to the bispectrum\footnote{We could reclaim these missing triangles by adding transverse shifts in addition to the radial shifts.}, we obtain all triangles that come from any pair of skewers.

\begin{figure}
    \centering
    \includegraphics[width=1\linewidth]{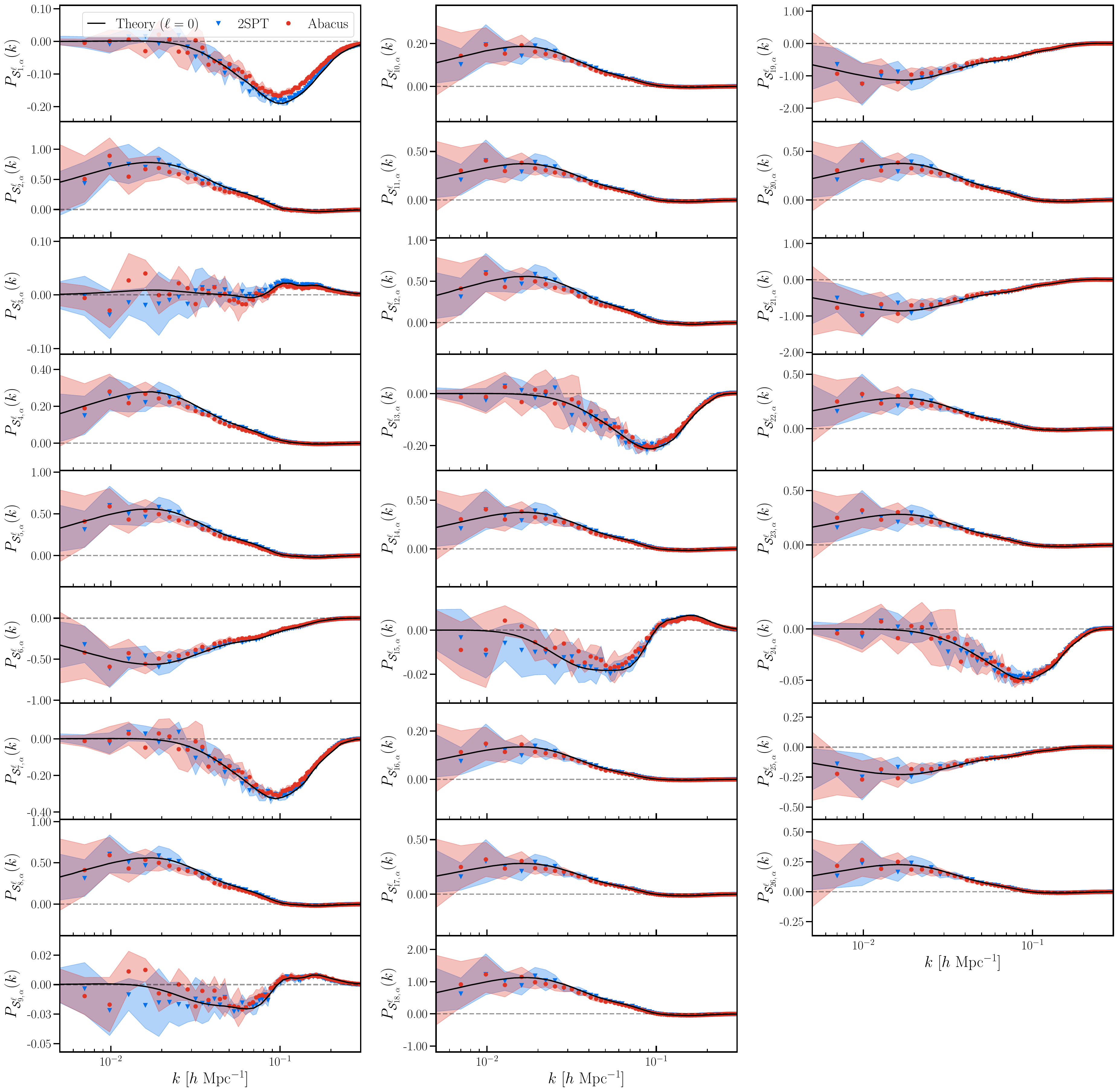}
    \caption{
    Same as Fig.~\ref{fig:comparison_2SPT_abacus_model3_ell0} for the monopole of the shifted skew spectra $\mathcal{S}_{n,\, \alpha}$, given in Eq.~\eqref{eqn:shifted_kernel}, using $\alpha=20 \hinvMpc$ and only applying line-of-sight smoothing with $R_\parallel=10 \hinvMpc$. 
   }
    \label{fig:comparison_2SPT_abacus_model3_ell0_shifted}
\end{figure}

\begin{figure}
    \centering
    \includegraphics[width=1\linewidth]{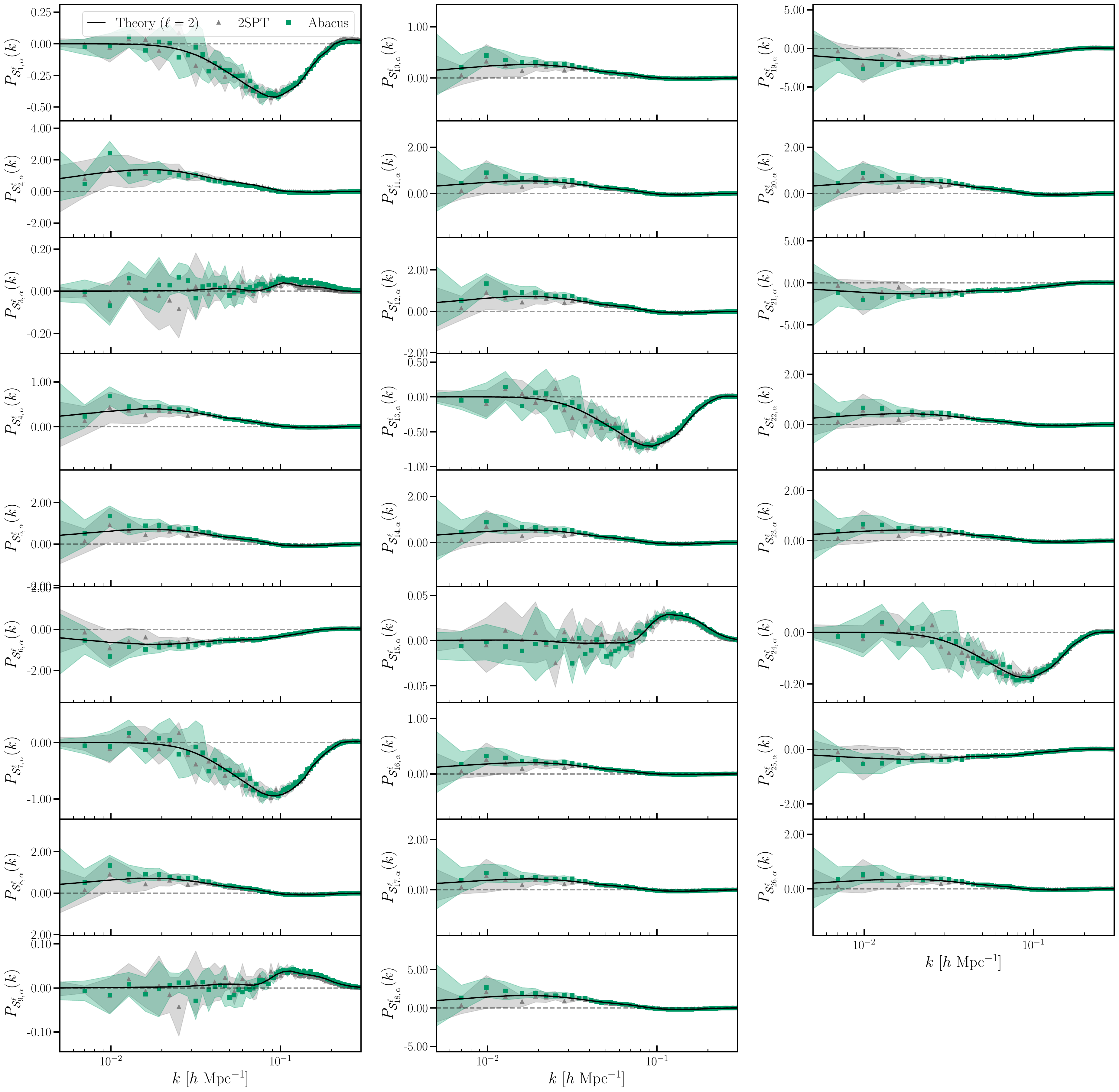}
    \caption{Same as Fig.~\ref{fig:comparison_2SPT_abacus_model3_ell0_shifted} for the quadrupole of the shifted skew spectra using $\alpha=20 \hinvMpc$ and only applying line-of-sight smoothing with $R_\parallel=10 \hinvMpc$.}
    \label{fig:comparison_2SPT_abacus_model3_ell2_shifted}
\end{figure}

The key validation results are shown in Figs.~\ref{fig:comparison_2SPT_abacus_model3_ell0_shifted}-\ref{fig:comparison_2SPT_abacus_model3_ell2_shifted}. We compare for both multipoles the shifted skew spectra where we use a fixed displacement between two points of $\alpha=40\hinvMpc$. The predicted theory curves (black solid lines) agree well with the measured skew spectra from synthetic perturbative fields (2SPT in blue) and from the \abacus simulations for the monopole and the quadrupole. By construction, we expect (and find) good agreement between the 2SPT fields and the theory on large scales which gives us confidence in the pipeline. 
Decreasing the anisotropic smoothing scale from $R_\parallel=10 \hinvMpc$ 
to $R_\parallel=5\hinvMpc$ yields a similar level of agreement as shown in App.~\ref{app:shifted_skew_R5}. Note that decreasing the line-of-smoothing scale down to $R_\parallel=1\hinvMpc$ results in discrepancies between the theory prediction and measured skew spectra indicating a breakdown of the tree-level EFT prescription. Similarly, for the baseline results when $\alpha$ approaches $R$ in magnitude we also get discrepancies. We leave to future work to directly fit for the skew spectra to constrain the bias parameters jointly with the one-loop power spectrum.

\subsection{Applicability to observational data: Forward modeling of the window matrix} \label{sec:p3d_delta2}
A caveat of the present approach is that it does not take into account the window function as we are computing grid-based FFTs which are affected by the sparse $\td$-function type sampling of the \LyaF transverse to the line-of-sight. In App.~\ref{app:theory} we list the different types of skew spectra and for the special case of the squared term (i.e., the correlation of $\td^2$ with $\td$) we forward model the window analytically in the following.  The window function of the \LyaF is non-trivial given the sparse sampling of the field transverse to and dense sampling along the line-of-sight.\footnote{We use the terms `window function' and `window matrix' interchangeably.}  
We follow the approach proposed in the context of the \Lya forest in Ref.~\cite{deBelsunce:2024knf} and use a pair-count spectral estimator in configuration space \cite{Philcox2020, Philcox:2021}. In the following we briefly summarize the approach and refer the reader to Ref.~\cite{deBelsunce:2024knf} for a fuller presentation. In a nutshell, we avoid grid-based fast Fourier transforms by weighting each pixel pair by $\exp(i\mat{k}\cdot \mat{r})$, for wave vector  $\mat{k}$ and pixel pair separation $\mat r$ and thus measures the anisotropic power spectrum, $P_{\ell}(k)$. It is obtained via Fourier transforming ($\mathcal{F}$) the two-point function of the data field $D(\mat x)=n(\mat x)w(\mat x)\delta(\mat x)$ for background density $n$, weights $w$, and flux decrement $\delta$\footnote{Note that $DD$ are two general data fields; here, $\td^2$ and $\td$.} 
\begin{equation} \label{eq:FFT_2PCF}
    P(\mat k) \equiv \mathcal{F}\left[ DD(\mat r) \right] = \int \mathrm{d}^3 \mat r\  DD(\mat r)e^{i\mat{k}\cdot \mat{r}} \,,
\end{equation}
with some normalization. In practice, this corresponds to computing a sum over each pixel pair, $i,j$, in each $k$-bin $a$:
\begin{equation} \label{eq:DD_pk}
    \hat{P}^a_\ell =\frac{1}{\overline{V(nw)^2}}\sum_{i}\sum_{j,\, i \neq j} w_iw_j \td_{F,i}\td_{F,j} A^{a}_\ell(\mat{r}_i,\mat{r}_j) W(|\mat{r}_i-\mat{r}_j|; R_0) \ ,
\end{equation}
which is normalized by the survey volume and removes `self-skewer' counts. This avoids (correlated) uncertainties stemming from continuum estimation and is an advantage of performing explicit pair counting in configuration space. For observed \Lya spectra, the weights $w$ and transmitted flux fractions $\td_F$ are obtained from the continuum fitting process (see, e.g.,~\cite{DESI_lya_2024}); here, we use the modes from the simulations as pixel values and model the selection function using weights of unity. The kernel $A^{a}_\ell$ captures the anisotropies and $\vk$-dependence and we additionally introduce a pair-count window function, $W(r;R_0)$, to speed-up the estimator, effectively removing pairs with separations beyond $R_0\geq 150\hinvMpc$ (given in equation 7 in Ref.~\cite{deBelsunce:2024knf}). The window matrix $\Phi(\rvec)\equiv DD(\rvec)/\xi(\rvec)$, for a data field $D({\mat x})$ is computed in configuration space and consists of the (weighted) mask and is  independent of the flux decrement, $\delta_F$,
\begin{equation} \label{eq:Phi-estimator}
    \Phi^b_\ell = \frac{1}{\overline{V(nw)^2}}\frac{2\ell+1}{V_b}\sum_{i}\sum_{j,\, i \neq j} w_iw_j\mathcal{L}_\ell(\hat{\mat r}_i\cdot\hat{\mat r}_j)\Theta^b(|\mat r_i-\mat r_j|) \, ,
\end{equation}
within some radial bin $b$ for Legendre multipole $\ell$. 

We convolve the theoretically expected power spectrum from Sec.~\ref{sec:results_III} with the window matrix instead of removing the window function from the data. This avoids numerical instabilities or binning the data in coarse pixels on the sky. Note that in contrast to galaxy surveys the \Lya forest does not require catalogs of random particles to compute the window. The pair-count estimator measures the power spectrum convolved with the survey geometry and the pair-count window $\Gamma_{\ell}^2(r) \equiv \Phi_{\ell}(r)W(r;R_0)$. The final survey geometry and pair-count-window convolved correlation function is given by (see, e.g.,~\cite{Castorina2018, Beutler:2016, 2021JCAP...11..031B}):
\begin{align} 
	\tilde{\xi}_{0}(r) = \xi_{0}\Gamma_{0}^2  &+ \frac{1}{5}\xi_2\Gamma^2_2 + \frac{1}{9}\xi_4\Gamma^2_4 + \dots
	\label{eq:conv1}\\
	\begin{split}
		\tilde{\xi}_{2}(r) = \xi_{0} \Gamma_{2}^2  &+ \xi_2\left[\Gamma^2_0 + \frac{2}{7}\Gamma^2_2 + \frac{2}{7}\Gamma^2_4\right]+\xi_4\left[\frac{2}{7}\Gamma^2_2 + \frac{100}{693}\Gamma^2_4 + \frac{25}{143}\Gamma^2_6\right] + \dots
		\label{eq:conv2}
	\end{split}
\end{align}
which we truncate beyond the hexadecapole for the theory correlation function while  including \textit{all} corresponding window matrix multipoles for each multipole in $\tilde{\xi_{\ell}}$. The window convolved correlation functions  are related to the power spectra through the Hankel transform. 

\begin{figure}
    \centering
    \includegraphics[width=0.49\linewidth]{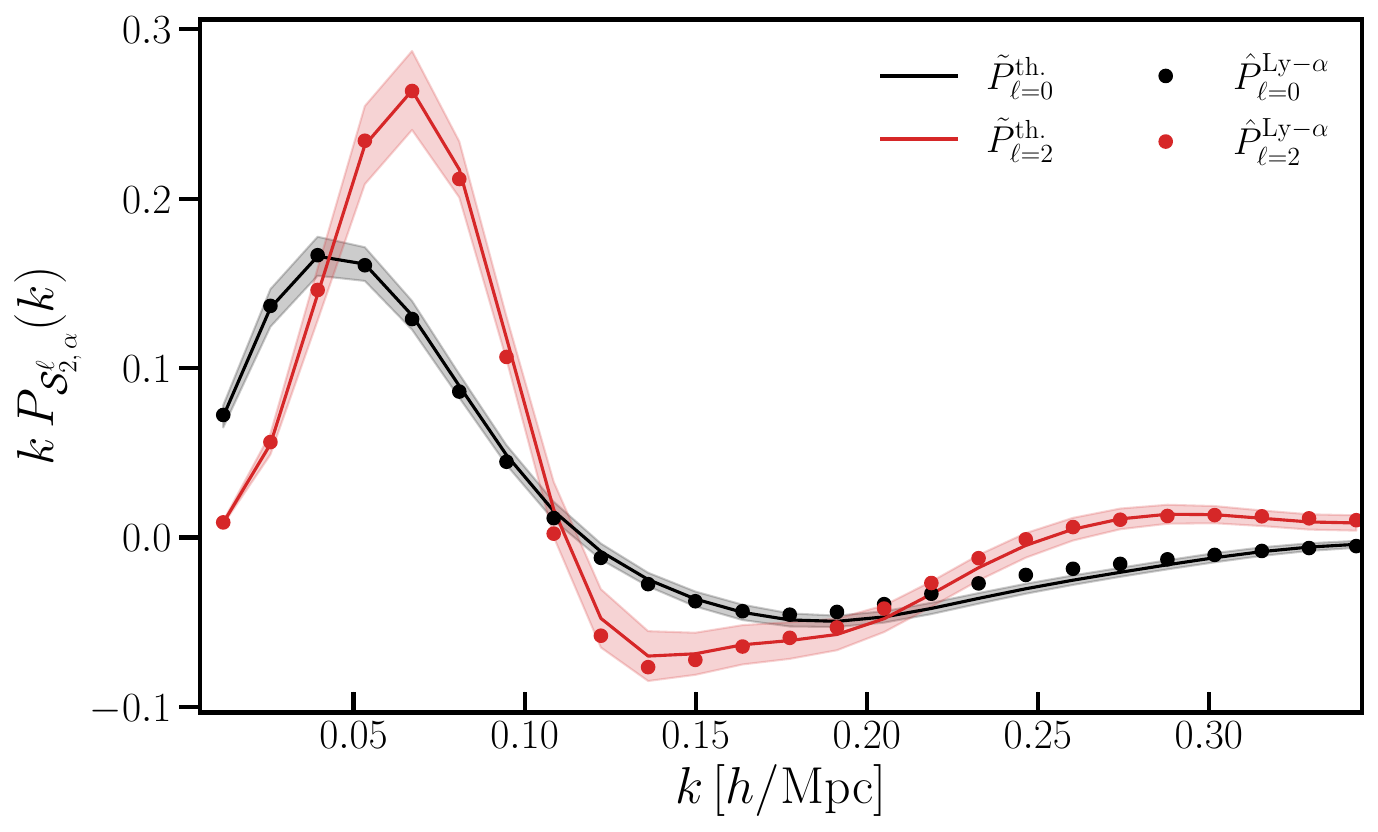}\hfill
    \includegraphics[width=0.49\linewidth]{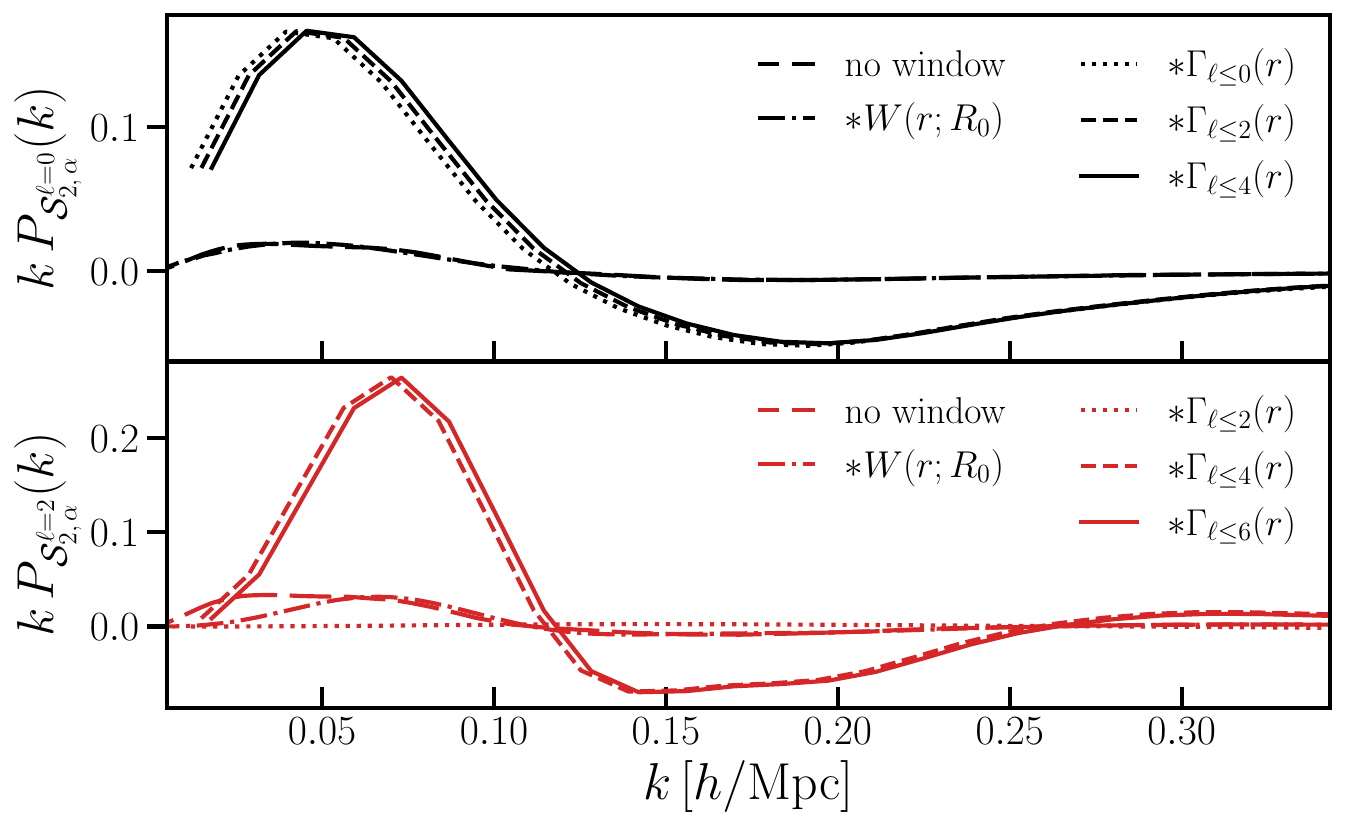}
    \caption{\textit{Left panel:} 
    Consistency test of the power spectrum estimator. We use the same fields as in Sec.~\ref{sec:results_III}, apply a survey geometry roughly emulating a sparse survey density similar to the one in eBOSS DR16 \cite{dMdB:2020} with 35 qso/deg$^2$ and compare the anisotropic window-convolved theory skew spectrum $\tilde{P}_{\mathcal{S}_{2,\,\alpha}^{\ell}}(k)$ (monopole in black and quadrupole in red) to the mean power spectrum, $\Hat{P}(k)$, of $N=65$ 2SPT simulations (dots). We apply a line-of-sight smoothing of $R_{\parallel}=10 \hMpcinv$, a shift parameter of $\alpha=20\hMpcinv$ and pair counts are truncated at $R_0=150\hMpcinv$ for computational efficiency. The shaded regions give the square root of the diagonal of the covariance matrix and illustrate the variance between the $N$ realizations.
    \textit{Right panel:} Illustration of the effect of the forward modeling of the window and truncation of the multipoles contributing to Eqs.~\eqref{eq:conv1}-\eqref{eq:conv2}. The solid lines are our final theory predictions that we compare to data in the left panel. To visualize the impact of adding higher order multipoles to the window modeling, we shift the dotted, dashed and solid lines along the $k$-axis by multiples of $\Delta k = 0.005$. 
    }  
    \label{fig:p3d_masked_skewspec}
\end{figure}

We compare the measured skew spectrum multipoles\footnote{We use the following notation: $\Hat{P}(k)$ is 3D power spectrum measured on the data. Theory spectra are denoted by $P(k)$ which, when convolved with the window matrix, are labeled $\Tilde{P}(k)$. The multipoles of the power spectra are denoted by the subscript $\ell$, \ie, $P_{\ell}(k)$.} to the theory input skew spectrum in the left panel of Fig.~\ref{fig:p3d_masked_skewspec} for $\mathcal{S}_{2,\,\alpha}$ where we correlate $\langle \delta_F(\vs')\delta_F(\vs-\Delta\vs)\delta_F(\vs+\Delta\vs)\rangle$ with $\Delta \vs(\alpha,\hat{z}) = \alpha\hat{z}$. By varying the shift parameter $\alpha$ we can probe configurations beyond the squeezed limit. We introduce a non-trivial window by sampling $N_{\rm s}=30,000$ sight lines from the box, corresponding to a quasar density of $\sim$\,$35~\mathrm{qso}/\mathrm{deg}^2$, and $14.5\cdot10^6$ \Lya pixels. We use a line-of-sight smoothing of $R=10\hMpcinv$ and a shift parameter of $\alpha=20 \hMpcinv$. The power spectrum is measured in bins of $\Delta k=0.014\hMpcinv$ in the range $(0.01 \leq k \leq 0.35)\hMpcinv$.  We recover the input spectrum within the $1\sigma$ error bars from $N=65$ realizations up to $k \leq 0.35 \hMpcinv$ for monopole (black) and quadrupole (red). The error bars shown in the left panel of Fig.~\ref{fig:p3d_masked_skewspec} capture the variation between $N$ realizations. Note that the quadrupole crossing the monopole at $k\sim 0.05\hinvMpc$ stems from the pair truncation window that is set to $R_0=150\hMpcinv$ for computational efficiency. Thus, information beyond $\pi/(R_0/2)$ is already suppressed by approximately a factor of two. We emphasize that the pair truncation window does not introduce any 3D smoothing; it just restricts the maximum separation $R_0$ between pixel pairs that are used when computing the power spectrum. 

In the right panel of Fig.~\ref{fig:p3d_masked_skewspec} we show the impact of forward modeling the window matrix into the theory prediction for the monopole in black (quadrupole in red) in the top (bottom) panel. We start from the theory prediction without the window function and gradually add complexity (the dotted, dashed and solid lines are shifted along the $k$-axis for illustration purposes): first, the pair truncation window, only altering large-scale power. Second, only keeping the monopole in the mode mixing terms given in Eqs.~\eqref{eq:conv1}-\eqref{eq:conv2}. Third, adding in the quadrupole and, finally, adding in the hexadecapole. It is interesting to note that for the monopole $\Gamma_{\ell=0}$ captures all the relevant information and for the quadrupole $\Gamma_{\ell=2}$ is required. For the quadrupole the maximum multipole that is folded into the theory prediction is adapted accordingly ($\ell\leq 6$).\footnote{We verified that adding higher order moments does not change the conclusion of this paper.} 

The excellent agreement between theory and measurement shows that this procedure is readily applicable to data and will help to break the mean flux degeneracy present in the power spectrum \cite{2001ApJ...551...48Z, 2003MNRAS.344..776M}. For the skew spectra involving derivatives of the window or inverse Laplacian terms, we advocate computing the expected windowed skew spectra using a Monte Carlo approach. A potential recipe would look like this: First, we use the tree-level EFT prescription to generate the \Lya fields at the field-level (2SPT fields). Second, we apply all the skew spectra given in Sec.~\ref{sec:skewspec_operators} to the two copies of the field. Then, we apply a realistic \LyaF survey geometry to both fields, \textit{i.e.}~the original flux decrement and to all quadratic operators arising in the bispectrum ($\mathcal{S}_1,\,\mathcal{S}_3$-$\mathcal{S}_{26}$). Finally, we measure the power spectrum using the above discussed weighted pair count estimator. The final result will be the window convolved skew spectrum $\tilde{P}_{\mathcal{S}^{\ell}_{n,\, \alpha}}$ which we can map to the unwindowed one. Now we repeatedly draw ($\simgt 10^5$ samples) from the priors of the bias parameters (which we assume to be independent of cosmology) yielding an emulator for the remaining 25 windowed skew spectra -- we leave this to future work.

\section{Summary and Conclusions} \label{sec:conclusions}
Measurements of the three-dimensional clustering of the \Lya forest are reaching sub-percent-level constraints on the cosmic expansion in the matter-dominated regime at redshift $z = 2.33$ \cite{DESI:2025zpo}, yielding complimentary information to low-redshift ($z<1$) galaxy surveys \cite{DESI:2025zpo}. Galaxy clustering measurements are increasingly including higher-order information \cite[see, e.g.,~][]{2022PhRvD.105d3517P} in contrast to \Lya forest analyses. The influx of medium-resolution \Lya forest spectra from the currently observing Dark Energy Spectroscopic Instrument \cite[DESI;][]{DESI:2016, DESI:2022}, paired with recent advances in the development of the one-loop EFT power spectrum \cite{Ivanov2024}, opens new opportunities to extract cosmological parameters from the \Lya forest. A limitation for the \Lya forest power spectrum is the degeneracy between then mean flux and power spectrum amplitude which can be broken using higher order statistics \cite{2001ApJ...551...48Z, 2003MNRAS.344..776M}. Directly measuring the three-point correlation function, however, is a challenging task: (i) the large number of spectral pixels makes it computationally very expensive; and (ii) the sparse $\delta$-function sampling of the \Lya forest transverse to the line-of-sight affects standard grid-based Fourier transforms (see, e.g.,~\cite{Font-Ribera:2018, deBelsunce:2024knf}).

In this work, we have presented the derivation of the tree-level bispectrum of the \Lya forest in the framework of the effective field theory of large-scale structure (EFT) directly in redshift space. 
We use the resulting second-order bias expansion to compute 26 skew spectra for the \Lya forest; twelve more than for galaxies in redshift space given the additional line-of-sight-dependent terms. 
These skew spectra, cross-spectra between the input density field and quadratic operators applied to this field, are expected to capture most of the information on 
bias parameters and the growth rate as the full bispectrum \cite{2015PhRvD..91d3530S,Schmittfull:2021}. 
We also introduce the \textit{shifted skew spectra}, which allow us to probe other shapes of the bispectrum beyond the squeezed limit probed by the usual skew spectra.
We validate our methodology on two sets of simulations: (i) synthetic three-dimensional 
\Lya fields in redshift-space using perturbation theory up to second order, and (ii) large \Lya forest mocks produced on the $N$-body simulation suite \textsc{AbacusSummit}. 
Using the tree-level bispectrum theory model, we find excellent agreement with the synthetic three-dimensional \Lya fields and agreement at the $1$-$2\sigma$ level up to $k \simlt 0.17 \hMpcinv$ for the \abacus simulations. 
We validate our methodology and pipeline using an idealized scenario with an isotropic smoothing function. 
To enable data analysis of observed spectra where we cannot 3D smooth the data we instead use an anisotropic smoothing along the line of sight - which we find produces good agreement between simulated and theoretical skew spectra.
We show that the (windowed) skew spectrum $\mathcal{S}_2$ is directly applicable to \LyaF data (e.g., from DESI) by forward modeling the window; yielding a similar agreement with the theory prediction up to $k \simlt 0.34 \hMpcinv$ for a radial displacement of $\alpha=20\hinvMpc$ and line-of-sight smoothing parameter of $R_{\parallel}=5\hinvMpc$. 

This work is the first step in extracting information from the three-dimensional bispectrum of the \Lya forest. The shifted skew spectra are readily applicable to simulations. For observational data (in the presence of a window function) we show how to forward model this exactly for the cross-spectrum of  the squared field with the original one and discuss the window function treatment for other skew spectra in Appendix~\ref{app:theory}. 
Further generalizing the shifted skew spectra including a transverse shift supplies an alternative method of probing the full bispectrum without explicitly modeling the terms that are proportional to derivatives of the window. 
This series of tests shows a promising path to extract more information from the \LyaF and should be used as motivation for using these cross-spectra as an additional summary statistic. 

There are several immediate directions for extension of the work presented here. 
Concrete practical algorithms
can be developed to estimate the skew spectra 
involving derivatives from real data, and make
theory predictions for the window-complicated
results. 
The numerical evaluation of the skew spectra is somewhat cumbersome but should be easily amenable to speed up with FFTLog techniques for the \Lya forest kernels \cite{Simonovic:2017mhp,Chudaykin:2020aoj, Ivanov2024}.
One of the most compelling targets for the \Lya forest is primordial non-Gaussianity -- the addition of the small number of additional requisite skew spectra is straightforward from the template shapes of the primordial bispectra \cite{MoradinezhadDizgah:2019xun,2024JCAP...05..011C}.
The presented theoretical formalism can also be readily applied to the \textit{cross}-correlation measurements of cosmic microwave background (CMB) lensing with the one-dimensional \Lya forest power spectrum, effectively probing a squeezed bispectrum \cite{Doux:2016, 2024PhRvD.110f3505K} or to multi-tracer skew spectra (e.g.,~\Lya\ - \Lya\ - quasar and \Lya\ - quasar - quasar requiring a shot noise correction \cite{Dizgah_2020}) which could be used to test the equivalence principle as the \Lya forest traces baryons and quasars (mostly) trace dark matter (see, e.g.,~\cite{Peloso:2013zw,Kehagias:2013yd}). 
These topics will be the subject of upcoming work.

\section*{Acknowledgments}
The authors thank Mikhail Ivanov, Zvonimir Vlah, Fabian Schmidt, Shu-Fan Chen, Martin White, Oliver Philcox, and An\v{z}e Slosar for valuable discussions. They are especially grateful to Azadeh Moradinezhad Dizgah and Stephen Chen for their numerous insightful exchanges and to Emanuele Castorina for bringing the issue of masked skew spectra to their attention. 

RB and PM are supported by Lawrence Berkeley National Laboratory and the Director, Office of Science, Office of High Energy Physics of the U.S. Department of Energy and this research used resources of the National Energy Research Scientific Computing Center (NERSC), a U.S. Department of Energy Office of Science User Facility operated under Contract No.~DE–AC02–05CH11231.  
JMS acknowledges that, in part, support for this work was provided by The Brinson Foundation through a Brinson Prize.

\appendix

\section{\label{app:kspace_skew} Fourier-space skew spectra}
\begingroup
\allowdisplaybreaks
\begin{align}
    b_{1}^{3}f^{0}&: \, \mathcal{S}_{1} = F_{2}(\vk_1,\vk_2)\\
    b_{1}^{2}b_{2}f^{0}&: \, \mathcal{S}_{2} = 1\\
    b_{1}^{2}b_{\mathcal{G}_2}f^{0}&: \, \mathcal{S}_{3} = \mathcal{G}_2(\vk_1,\vk_2)\\
    b_{1}^{2}\textcolor{blue}{b_{(KK)_\parallel}}f^{0}&: \, \mathcal{S}_{4} = (KK)_{\parallel}(\vk_1,\vk_2)\\
    b_{1}^{2}\textcolor{blue}{b_{\Pi^{[2]}_\parallel}}f^{0}&: \, \mathcal{S}_{5} = \Pi^{[2]}_{\parallel}(\vk_1,\vk_2)\\
    b_{1}^{3}f^{1}&: \, \mathcal{S}_{6} = \frac{k_{1\parallel} k_{2\parallel}}{k_{1}^{2}} + \frac{k_{1\parallel} k_{2\parallel}}{k_{2}^{2}} \\
    b_{1}^{2}\textcolor{red}{b_{\eta}}f^{1}&: \, \mathcal{S}_{7} =  \left(\frac{k_{1\parallel}^{2}}{k_{1}^{2}} + \frac{k_{2\parallel}^{2}}{k_{2}^{2}}  \right)F_2(\vk_1,\vk_2) +\frac{k_{3\parallel}^{2}}{k_{3}^{2}}G_{2}(\vk_1,\vk_2)\\
    b_{1}b_{2}\textcolor{red}{b_{\eta}}f^{1}&: \, \mathcal{S}_{8} = \frac12 \left(\frac{k_{1\parallel}^{2}}{k_{1}^{2}} 
+ \frac{k_{2\parallel}^{2}}{k_{2}^{2}}  \right)\\
    b_{1}b_{\mathcal{G}_2}\textcolor{red}{b_{\eta}}f^{1}&: \, \mathcal{S}_{9} = \frac12 \left(\frac{k_{1\parallel}^{2}}{k_{1}^{2}} + \frac{k_{2\parallel}^{2}}{k_{2}^{2}}  \right)\mathcal{G}_{2}(\vk_1,\vk_2)\\
    b_{1}\textcolor{red}{b_{\eta}}\textcolor{blue}{b_{(KK)_\parallel}}f^{1}&: \, \mathcal{S}_{10} = \frac12 \left(\frac{k_{1\parallel}^{2}}{k_{1}^{2}} + \frac{k_{2\parallel}^{2}}{k_{2}^{2}}  \right)(KK)_{\parallel}(\vk_1,\vk_2) \\
    b_{1}\textcolor{red}{b_{\eta}}\textcolor{blue}{b_{\Pi^{[2]}_\parallel}}f^{1}&: \, \mathcal{S}_{11} = \frac12 \left(\frac{k_{1\parallel}^{2}}{k_{1}^{2}} + \frac{k_{2\parallel}^{2}}{k_{2}^{2}}  \right)\Pi_{\parallel}^{[2]}(\vk_1,\vk_2)\\
    b_{1}^{2}\textcolor{red}{b_{\delta\eta}}f^{1}&: \, \mathcal{S}_{12} = \frac12 \left(\frac{k_{1\parallel}^{2}}{k_{1}^{2}} + \frac{k_{2\parallel}^{2}}{k_{2}^{2}}  \right)  \\
    b_{1}(\textcolor{red}{b_{\eta}})^2 f^{2}&: \, \mathcal{S}_{13} = \frac{k_{1\parallel}^{2}k_{2\parallel}^{2}}{k_{1}^{2}k_{2}^{2}}  F_2(\vk_1,\vk_2) +   \left(\frac{k_{1\parallel}^{2}}{k_{1}^{2}} + \frac{k_{2\parallel}^{2}}{k_{2}^{2}} \right)\frac{k_{3\parallel}^{2}}{k_{3}^{2}}G_{2}(\vk_1,\vk_2)\\
    b_{2}(\textcolor{red}{b_{\eta}})^2 f^{2}&: \, \mathcal{S}_{14} = \frac{k_{1\parallel}^{2}k_{2\parallel}^{2}}{k_{1}^{2}k_{2}^{2}}\\
    b_{\mathcal{G}_2}(\textcolor{red}{b_{\eta}})^2 f^{2}&: \, \mathcal{S}_{15} = \frac{k_{1\parallel}^{2}k_{2\parallel}^{2}}{k_{1}^{2}k_{2}^{2}}\mathcal{G}_2(\vk_1,\vk_2)\\
    (\textcolor{red}{b_{\eta}})^2 \textcolor{blue}{b_{(KK)_\parallel}} f^{2}&: \, \mathcal{S}_{16} = \frac{k_{1\parallel}^{2}k_{2\parallel}^{2}}{k_{1}^{2}k_{2}^{2}}(KK)_{\parallel}(\vk_1,\vk_2)\\
    (\textcolor{red}{b_{\eta}})^2 \textcolor{blue}{b_{\Pi^{[2]}_\parallel}} f^{2}&: \, \mathcal{S}_{17} = \frac{k_{1\parallel}^{2}k_{2\parallel}^{2}}{k_{1}^{2}k_{2}^{2}}\Pi_{\parallel}^{[2]}(\vk_1,\vk_2)\\
    b_{1}\textcolor{red}{b_{\eta}}\textcolor{red}{b_{\delta \eta}}f^{2}&: \, \mathcal{S}_{18} = 
    \frac12 \left(\frac{k_{1\parallel}^{4}}{k_{1}^{4}} + \frac{k_{2\parallel}^{4}}{k_{2}^{4}}\right) + 2\frac{k_{1\parallel}^2 k_{2\parallel}^{2}}{k_1^2 k_{2}^{2}} \\
    b_{1}^{2}\textcolor{red}{b_{\eta}}f^{2}&: \, \mathcal{S}_{19} = \frac12 \left( \frac{k_{1\parallel}^{3}k_{2\parallel}}{k_{1}^{4}} + \frac{k_{1\parallel}k_{2\parallel}^{3}}{k_{2}^{4}} + 2 \frac{k_{1\parallel}^{3}k_{2\parallel}+k_{1\parallel}k_{2\parallel}^{3}}{k_{1}^{2} k_{2}^{2}}\right) \\
    b_{1}^{2}\textcolor{red}{b_{\eta^2}} f^{2}&: \, \mathcal{S}_{20} = \frac{k_{1\parallel}^{2}k_{2\parallel}^{2}}{k_{1}^{2} k_{2}^{2}}\\
    b_1 (\textcolor{red}{b_{\eta}})^2f^{3}&: \, \mathcal{S}_{21} =  \frac12\left( \frac{k_{1\parallel}^{5}k_{2\parallel}}{k_1^4 k_{2}^{2}}  + \frac{k_{1\parallel}k_{2\parallel}^5}{k_{1}^{2} k_{2}^{4}}  + 2 \left( \frac{k_{1\parallel}^{3}k_{2\parallel}^{3}}{k_{1}^{2} k_{2}^{4}} + \frac{k_{1\parallel}^{3}k_{2\parallel}^{3}}{k_{1}^{4} k_{2}^{2}}\right) \right)\\
    b_1 \textcolor{red}{b_{\eta} b_{\eta^{2}}}f^{3}&: \, \mathcal{S}_{22} = \frac12\left( \frac{k_{1\parallel}^{4}k_{2\parallel}^{2}}{k_{1}^{4} k_{2}^{2}} + \frac{k_{1\parallel}^{2}k_{2\parallel}^{4}}{k_{1}^{2} k_{2}^{4}} \right) \\
    (\textcolor{red}{b_{\eta}})^2 \textcolor{red}{b_{\delta\eta}}f^{3}&: \, \mathcal{S}_{23} = \frac12 \left( \frac{k_{1\parallel}^{4}k_{2\parallel}^{2}}{k_{1}^{4} k_{2}^{2}} +\frac{k_{1\parallel}^{2}k_{2\parallel}^{4}}{k_{1}^{2} k_{2}^{4}} \right) \\
    (\textcolor{red}{b_{\eta}})^3f^{3}&: \, \mathcal{S}_{24} = \frac{k_{1\parallel}^{2} k_{2\parallel}^{2}}{k_{1}^{2} k_{2}^{2}} \frac{k_{3\parallel}^{2}}{k_{3}^{2}}G_{2}(\vk_1,\vk_2)\\
    (\textcolor{red}{b_{\eta}})^2 \textcolor{red}{b_{\eta^2}}f^{4}&: \, \mathcal{S}_{25} = \frac12 \left( \frac{k_{1\parallel}^{3}k_{2\parallel}^{5}}{k_{1}^{4} k_{2}^{4}} + \frac{k_{1\parallel}^{5}k_{2\parallel}^{3}}{k_{1}^{4} k_{2}^{4}} \right) \\
    (\textcolor{red}{b_{\eta}})^3f^{4}&: \, \mathcal{S}_{26} = \frac{k_{1\parallel}^{4}k_{2\parallel}^{4}}{k_{1}^{4} k_{2}^{4}}.
\end{align}
\endgroup

\section{Correlation matrix \label{app:corr_mat}}
\begin{figure}
    \centering
    \includegraphics[width=1\linewidth]{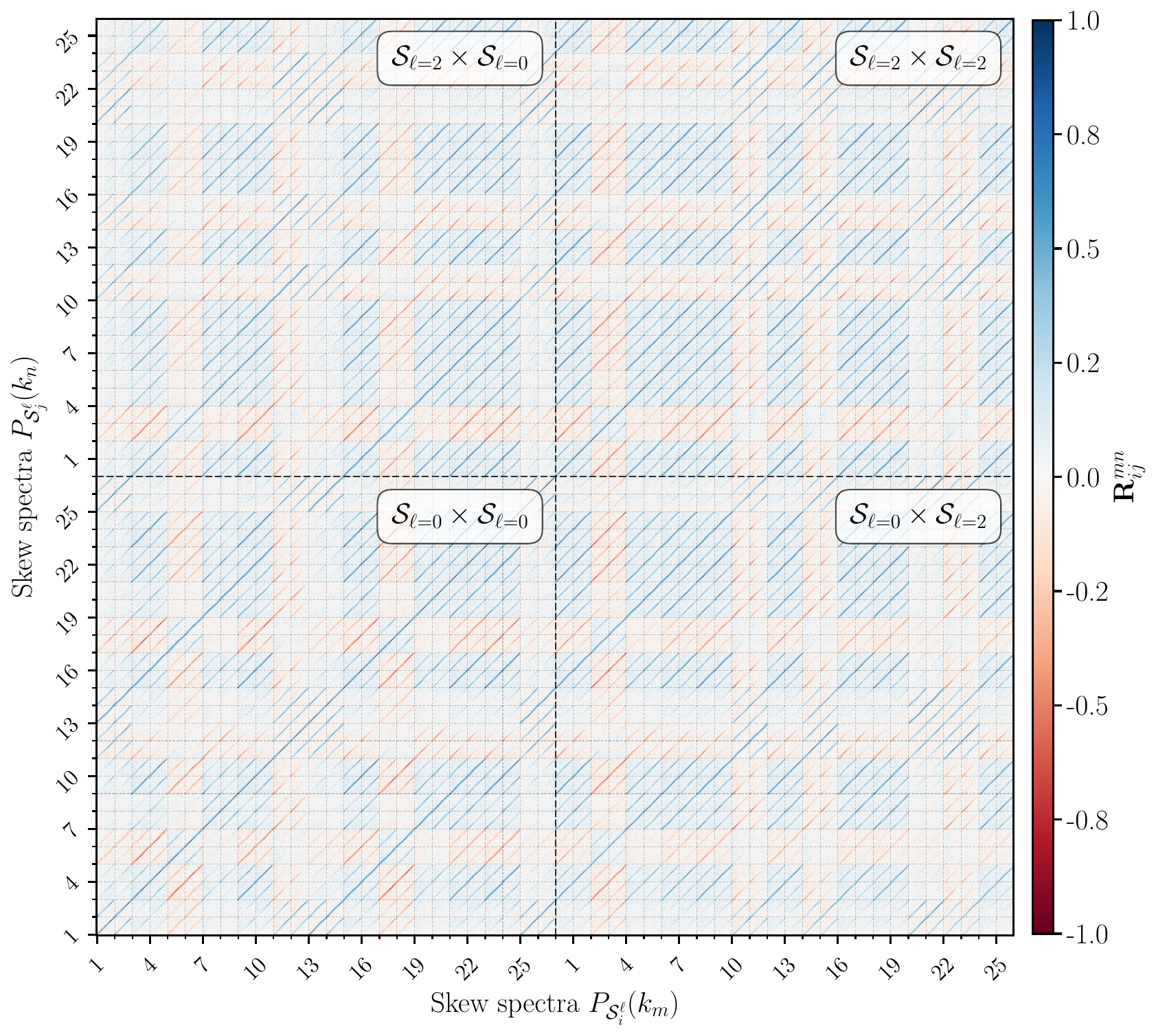}
    \caption{Correlation matrix of the skew spectrum multipoles $P_{\mathcal{S}^{\ell}_n}(k)$ measured from $N=1,000$ 2SPT field realizations using model III as bias parameter inputs. The correlation matrix $\mathbf{R}_{ij}^{mn}$ is defined in Eq.~\eqref{eq:corr_mat}; red denotes fully correlated and blue fully anti-correlated bins.}
    \label{fig:corr_mat}
\end{figure}

The error bars shown in Fig.~\ref{fig:comparison_2SPT_abacus_model3_ell0} capture the variation between differing realizations of the 2SPT field. For ease of comparison, we compute these with the same number as available \abacus simulations. To illustrate how correlated the different skew spectra are, we compute $N=1,000$ skew spectra of different 2SPT fields. For each skew spectrum monopole $i, j$ of the power in $k$-bins $m, n$, the covariance matrix is given by 
\beq \mathbf{C}_{\ell_1\ell_2}^{mn} = \frac{1}{N-1} \sum_{i=1}^N \left(\hat{S}^{m}_{\ell_1,i}-\overline{S}^{m}_{\ell_1}\right)\left(\hat{S}^{n}_{\ell_2,i}-\overline{S}^{n}_{\ell_2}\right) \,, \,  \text{with mean\, } \overline{S}^{m}_{\ell} = \frac{1}{N}\sum_{i=1}^N\hat{S}^{m}_{\ell,i}\,, \eeq 
where 
$i$ indexes the $N$ simulations. The correlation matrix is defined as
\beq \mathbf{R}^{mn}_{\ell_1\ell_2} \equiv \frac{\mathbf{C}_{\ell_1\ell_2}^{mn}}{\sqrt{\mathbf{C}_{\ell_1\ell_1}^{mm}\mathbf{C}_{\ell_2\ell_2}^{nn}}}\,, \label{eq:corr_mat}\eeq
by construction, this is unity along the diagonal. We show all inter- and intra-multipole combinations of the correlation matrix in Fig.~\ref{fig:corr_mat}. The data vector is $S^{m}_{\ell_1},S^{m}_{\ell_2}$ resulting in four blocks for the correlation matrix. Within each $S^{m}_{\ell_1} \times S^{m}_{\ell_2}$ block columns $i$ and rows $j$ display the correlation between $k$-bins. Qualitatively, it is interesting to note that the covariance matrix is block-diagonal for each inter- and intra-multipole blocks, with small ($\simlt 10-20\%$) off-diagonal correlations.  We apply a scale cut to the Fourier wavenumber in the range $(0.05 \leq k \leq 0.15)\hMpcinv$ resulting in a data vector of length $N_{\text{k}} = 832$ $k$-bins. The covariance matrix should be taken as indicatively as for the number of data points one would require a large Hartlap correction factor \cite{2007A&A...464..399H} of $h=0.17$ to debias the inverse of the covariance matrix $\mathbf{C}^{-1} \rightarrow \left( N - N_{\text{k}} - 2 \right) / \left( N - 1 \right) \times \mathbf{C}^{-1}$. To fit the bias parameters either an analytic Gaussian covariance matrix or an order of magnitude more realizations is required resulting in  $h\approx 0.92$.\footnote{Note that we use the Hartlap correction factor as an approximate benchmark to assess the number of required realizations for a simulations-based covariance matrix.} 

\section{Smaller scale shifted skew spectra} \label{app:shifted_skew_R5}
\begin{figure}
    \centering
    \includegraphics[width=1\linewidth]{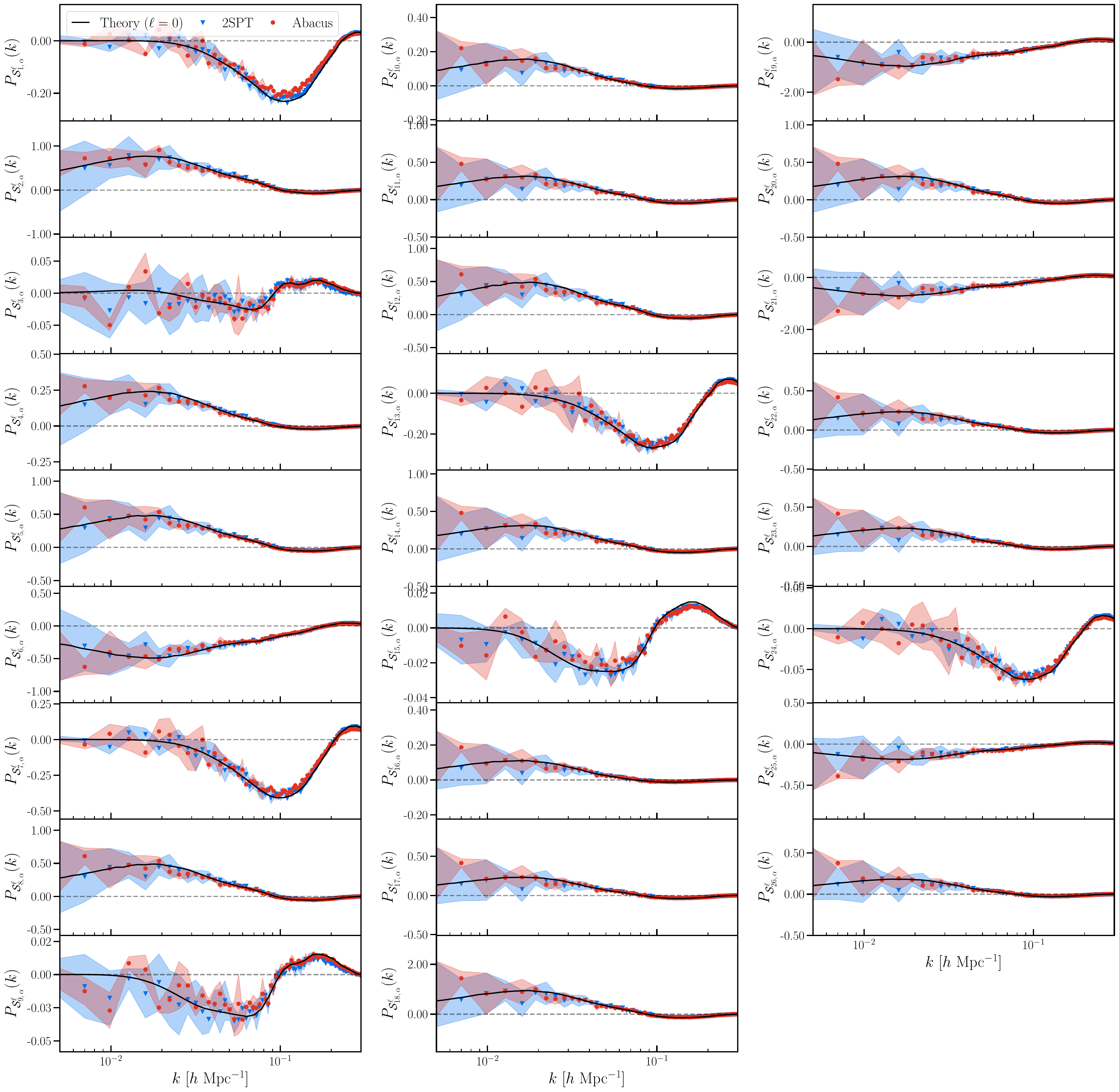}
    \caption{
    Same as Fig.~\ref{fig:comparison_2SPT_abacus_model3_ell0_shifted} for the monopole of the shifted skew spectra $\mathcal{S}_{n,\, \alpha}$, given in Eq.~\eqref{eqn:shifted_kernel}, using $\alpha=20 \hinvMpc$ and $R=5 \hinvMpc$. 
   }
    \label{fig:comparison_2SPT_abacus_model3_ell0_shifted_R5}
\end{figure}

\begin{figure}
    \centering
    \includegraphics[width=1\linewidth]{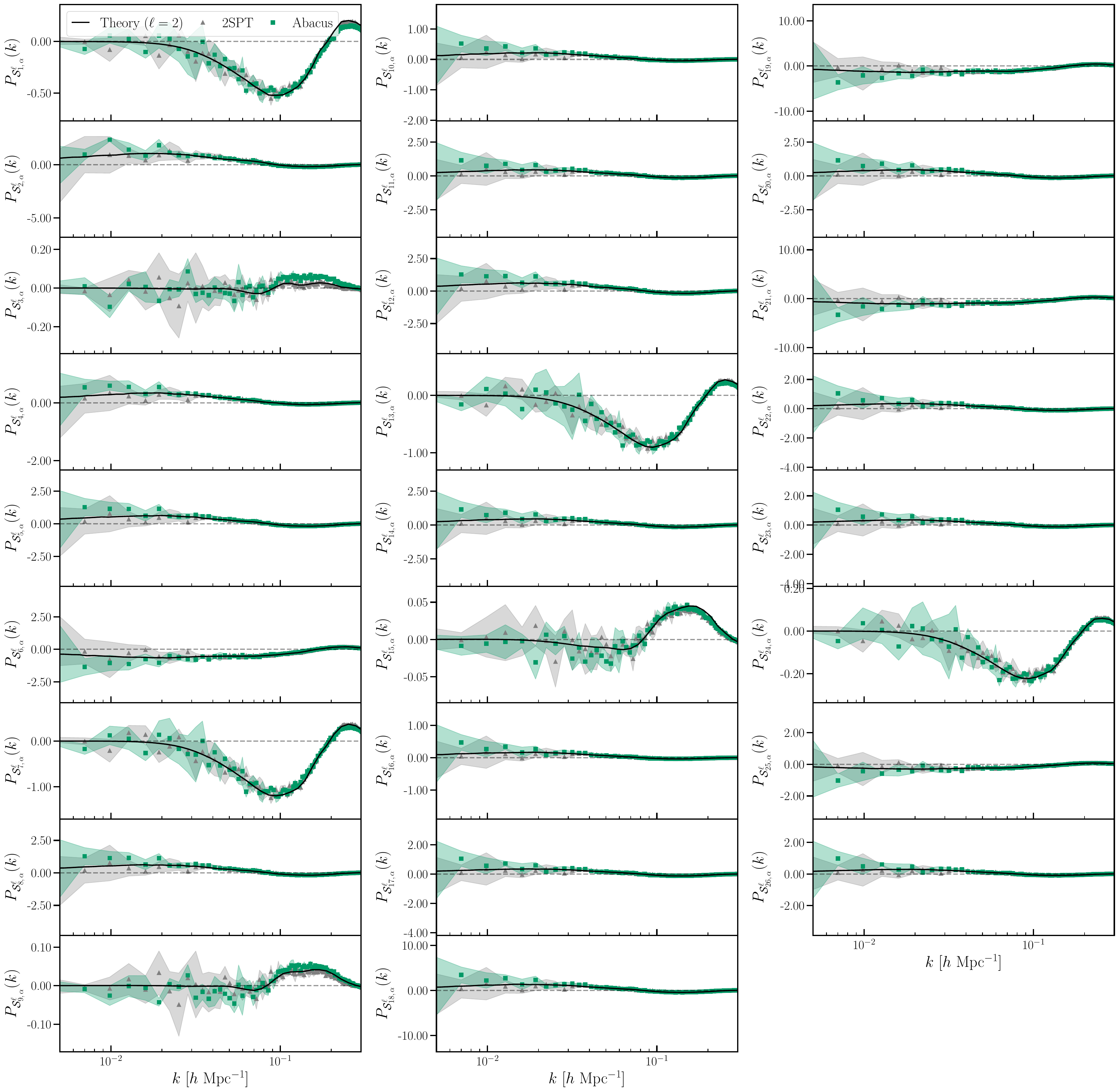}
    \caption{Same as Fig.~\ref{fig:comparison_2SPT_abacus_model3_ell0_shifted_R5} for the quadrupole of the shifted skew spectra using $\alpha=20 \hinvMpc$ and only applying line-of-sight smoothing with $R=5 \hinvMpc$.}
    \label{fig:comparison_2SPT_abacus_model3_ell2_shifted_R5}
\end{figure}

In this appendix we include smaller scales in the skew spectra by using $R_{\parallel}=5\hinvMpc$ and $\alpha=20 \hinvMpc$ as shown in Figs.~\ref{fig:comparison_2SPT_abacus_model3_ell0_shifted_R5}  and \ref{fig:comparison_2SPT_abacus_model3_ell2_shifted_R5} for the monopole and quadrupole, respectively. Decreasing the smoothing scale yields a further $k$-reach and an increase in amplitude and signal-to-noise ratio of the skew spectra. 

\section{Lyman-$\alpha$ Forest Skew Spectra and Survey Geometry\label{app:theory} } 

In this Appendix, we explore the form of the skew spectra in the presence of a survey geometry. 
In the context of the \Lya forest, this corresponds to a $\td$-function type sampling, \emph{i.e.} a dense (sparse) sampling along (transverse to) the line-of-sight. 
In Sec.~\ref{sec:p3d_delta2}, we measure the skew spectra of the \Lya forest flux fluctuations using the 3D power spectrum estimator  presented in \cite{deBelsunce:2024knf} applied to the product of a locally squared, $\td_F\td_F$, (where both fields are displaced by $\pm \alpha$) and the original field, $\td_F$.

For the flux power spectrum estimator, we have the familiar expression
\begin{equation}
    \Hat{P}^{(o)}_{\td_F\td_F} (k) = \int \frac{\diff \Omega_k}{4\pi} \int \diff^3\, \vx_1 \int \diff^3\, \vx_2 \, e^{-i\vk \cdot(\vx_2-\vx_1)} \td_F(\vx_1) \td_F(\vx_2)W(\vx_1)W(\vx_2) \ .
\end{equation}
For skew spectra, we are interested in the cross-spectra of one \LyaF field with a general quadratic operator $\mathcal{O}[X,Y](\vx)$ for two fields $X,Y$, which will be either $\td_F$ or some linear operator applied to it.
We can rewrite this as
\begin{equation} \label{eq:Pobs}
    \Hat{P}^{(o)}_{\mathcal{O}[\td_F,\td_F]\td_F} (k) = \int \frac{\diff \Omega_k}{4\pi} \int \diff^3\, \vx_1 \int \diff^3\, \vx_2 \, e^{-i\vk \cdot(\vx_2-\vx_1)} \mathcal{O}\left[ W\td_F,W\td_F \right](\vx_1) \td_F(\vx_2) W(\vx_2) \ ,
\end{equation}
where the \Lya forest signal is encoded in $\td_F(\vx)=\td_F(\chi\hat{\vx})$ (for quasar position $\Hat{\vx}$ and comoving distance $\chi$) with the corresponding mask, $W(\vx, \chi)$, which is given as a series of $\td$-functions $W(\hat{\vx},\chi) \equiv \sum_j \delta(\hat{\vx}-\hat{\vx}_j) W_j(\chi)$ for a sum over quasars. 

Unfortunately, unlike simple polynomials of the fields, nearly all of the skew spectrum operators $\mathcal{O}$ involve derivatives of the configuration space product of the window and the density field.
This means the resulting skew-spectra cross-power estimator means involve correlators in configuration space that are not separable with respect two powers of the window function $W$.
As always, one can work in Fourier space instead, at which point everything becomes separable.
However, this involves Fourier transforms of the window function itself, which, as mentioned, is very sparse for current \LyaF data and will continue to be so in the near future.
This leads to numerical issues when computing grid-based Fourier transforms, fundamentally limiting the practical application of the skew spectra to the Lyman-$\alpha$ forest.
In this work, we will show several representative examples of statistics that must be computed to account for the presence of the window.
We will reduce all expressions down to terms involving this fundamentally non-separable piece 
\begin{equation}
    \mathbf{v}[X;W](\vx) \equiv \frac{\nabla (WX)}{\nabla^2}. 
\end{equation}
It will also be convenient to consider two related functions $f,g$
\begin{equation}
    f[X,Y;W](\vx) \equiv \mathbf{v}[X;W](\vx) \cdot \nabla (W Y)(\vx)
\end{equation}
\begin{align}
    g[X,Y;W](\vx) &\equiv \left(\nabla \mathbf{v}[X;W](\vx)\right) \left(\nabla \mathbf{v}[Y;W](\vx)\right)\\
    &= \left(\partial_i v_j[X;W](\vx)\right) \left(\partial^i v^j [Y;W](\vx)\right)\nonumber
\end{align}
deferring the full treatment of this key issue to future work\footnote{We thank A. Moradinezhad for sharing notes and calculations on the window convolution, which informed our treatment of the derivative operators here; a related calculation for the galaxy skew spectra is in preparation.}. 

However, we give an indication of the challenge of this problem by considering the straightforward squared term as well as the first non-trivial example - the shift term that arises in the  $F_2$ quadratic operator.

The first skew spectrum is given by 
\begin{align}
\tilde{\mathcal{S}}_1 &= F_2[W \td_F,W \td_F](\vx) \,, \nonumber \\
&= \left[W(\vx)\td_F(\vx)\right]^2 - \Psi_{WF}(\vx) \cdot \nabla \left(W(\vx)\td_F(\vx)\right) + \frac{2}{7}\left(\left[\frac{\partial_i\partial_j}{\nabla^2} \left(W(\vx) \td_F(\vx)\right)\right]^2 - \left(W(\vx) \td_F(\vx)\right)^2  \right) \,,
\end{align}
where the $\tilde{\mathcal{S}}$ denotes the masked skew spectrum and $\Psi_{WF}(\vx) = \frac{\nabla}{\nabla^2}\left(W(\vx) \delta_F(\vx) \right)$. 
We derive the squared term and write the shift shift term below.

\subsubsection{Squared term}

For the squared term, we are able to write the expectation value skew spectrum estimator without any issues, as there are no derivative terms.
The ensemble average $\langle \dots \rangle$ of the squared term contribution is given by
\begin{align} \label{eq:Pobs_int}
    \langle \Hat{P}^{(o)}_{\td_F^2\td_F} (k) \rangle &= \int \frac{\diff \Omega_k}{4\pi} \int \diff^3 \vx_1 \int \diff^3 \vx_2\, e^{-i\vk \cdot(\vx_2-\vx_1)} \langle \td_F^2(\vx_1) \td_F(\vx_2) \rangle W^2(\vx_1)W(\vx_2)  \\
    &= \int \frac{\diff \Omega_k}{4\pi} \int \diff^3 \vx_1 \int \diff^3 \vx_{21}\, e^{-i\vk \cdot \vx_{21}} \sum_{L}\xi_{\td_F^2\td_F}^{(L)}(x_{21}) \mathcal{L}_{L}(\Hat{\vx}_{21}\cdot \Hat{\vx}_{1}) W^2(\vx_1)W(\vx_2)\\
    &= \int \diff^3 \vx_1 \int \diff^3 \vx_{21} j_0(k x_{21}) \sum_{L}\xi_{\td_F^2\td_F}^{(L)}(x_{21}) \mathcal{L}_{L}(\Hat{\vx}_{21}\cdot \Hat{\vx}_{1}) W^2(\vx_1)W(\vx_{21}+\vx_1)\\
    &= \int \diff x_{21} x_{21}^2  j_0(k x_{21}) \sum_{L}\xi_{\td_F^2\td_F}^{(L)}(x_{21}) \int \diff \Omega_{21} \int \diff^3 \vx_1 W^2(\vx_1)W(\vx_{21}+\vx_1) \mathcal{L}_{L}(\Hat{\vx}_{21}\cdot \Hat{\vx}_{1})\,,
\end{align}
where $x_{21}=|\vx_{21}|=|\vx_{2}-\vx_{1}|$ and $\langle \td_F^2(\vx_1) \td_F(\vx_2) W^2(\vx_1)W(\vx_2)\rangle = \langle \td_F^2(\vx_1) \td_F(\vx_2) \rangle W^2(\vx_1)W(\vx_2)$.
In the second equality we expand the correlation function in multipoles $L$ as $\xi(\vx_{12}) = \sum_{L} \xi^{(L)}(x_{12}) \mathcal{L}_{L}(\cos(\theta_{\vx_{12}}))$, work in the plane-parallel approximation, and employ the usual definitions of the spherical Bessel function of $0$-th order, $j_0$, and the Legendre polynomials $\mathcal{L}_L(\mu)$. We can re-write this in terms of the power spectrum as 
\begin{align}
    \langle \Hat{P}^{(o)}_{\td_F^2\td_F} (k) \rangle &= \sum_L (i)^L \int \frac{\diff p}{2\pi^2} p^2 P^{(L)}_{\td_F^2\td_F} (p) \int \diff x_{21} x_{21}^2 j_L(p x_{21})    j_0(k x_{21}) W_2^L(x_{21})\,,
\end{align}
with
\begin{equation}
    W_{2}^L(x_{21}) = \int \diff \Omega_{21} \int \diff^3 \vx_1 W^2(\vx_1)W(\vx_{21}+\vx_1) \mathcal{L}_{L}(\Hat{\vx}_{21}\cdot \Hat{\vx}_{1}) \,.
\end{equation}
which can be (almost) exactly estimated via pair counting \cite{deBelsunce:2024knf}.
This window can additionally include a pair truncation window $W(r;R_0)$ to reduce the computational burden of the estimator.
The normalization for the \Lya forest is, in principle, arbitrary; here, we  compute it via exhaustive pair counting over the weights of all three masks $w^2_i w_j$ for pixels $i$ and $j$.
The corresponding numerical results are shown in Sec.~\ref{sec:p3d_delta2}.

\subsubsection{Shift term}

The shift term is the product of the ``flux displacement'' field, $\Psi_F$, and the derivative of the flux field, $\partial_i\td_F$:
\begin{align}
    \label{eqn:shift}
\left[ \Psi_{WF}^i(\vx) \partial_i \left( W(\vx) \td_F(\vx) \right) \right] &=\left[ \frac{\partial_i}{\nabla^2} W(\vx) \td_F(\vx) \right] \partial_i \left( W(\vx) \td_F(\vx) \right)\,,\\
&= f[\td_{F}, \td_{F}; W](\vx)\nonumber
\end{align} 
which we plug into Eq.~\eqref{eq:Pobs}. 
\begin{align}
P_{[\Psi_F\partial_i\td_F]}(k) &= \int \diff^3\vx_1 \int \diff^3\vx_2\, e^{-i \vk \cdot (\vx_2 - \vx_1)} \left[-\frac{\partial_i}{\nabla^2} \left( W(\vx_1)\td_F(\vx_1) \right) \right] \partial_i \left[ W(\vx_1) \td_F(\vx_1) \right] W(\vx_2)\td_F(\vx_2)\,,\\
&= -\int \diff^3\vx_1 \int \diff^3\vx_2\, e^{-i \vk \cdot (\vx_2 - \vx_1)} f[\td_F,\td_F;W]  W(\vx_2)\td_F(\vx_2)\,.
\end{align} 
As already mentioned, $f$ is challenging to compute numerically via FFTs.
However, for completeness, we show the result that would proceed from attempting to compute this skew spectrum via FFTs.
Chain rule on the first two factors of the integrand expands to:
\beq
\underbrace{\left( \frac{(\partial_i W(\vx_1)) \td(\vx_i)}{\nabla^2}  \right) \left( \partial_i W(\vx_1) \right) \td_F(\vx_1)}_{\star_1}\,, 
\underbrace{\left( \frac{W(\vx_1) (\partial_i \td(\vx_i))}{\nabla^2}  \right) W(\vx_1) \left( \partial_i  \td_F(\vx_1)\right) }_{\star_2}\,, 
\eeq
and 
\beq
\underbrace{\left( \frac{(\partial_i W(\vx_1)) \td(\vx_i)}{\nabla^2}  \right) W(\vx_1) \left( \partial_i  \td_F(\vx_1)\right)}_{\star_3}\, ,
\underbrace{\left( \frac{W(\vx_1) (\partial_i \td(\vx_i))}{\nabla^2}  \right)  \left( \partial_i W(\vx_1) \right) \td_F(\vx_1)}_{\star_4}\, .
\eeq
The expectation value of each of these terms can be computed in Fourier space as
\begin{align}
    \star 1 &= \int_{\vk} e^{i \vk \cdot \vx} \nonumber \int_{\vk_1}\int_{\vk_2}\int_{\vq_1}\int_{\vq_2}\int_{\vp_1}\int_{\vp_2} \delta^{\rm D}(\vk+\vk_1+\vk_2)\delta^{\rm D}(\vk_1+\vq_1+\vq_2)\delta^{\rm D}(\vk_2+\vp_1+\vp_2)\big[\\
    &\quad\quad\quad\quad \frac{\vq_1 \cdot \vp_1}{k^2} W(\vq_1)  W(\vp_1)  \td_F(\vq_2) \td_F(\vp_2) 
    \big]\\ \nonumber
    \star 2 &= \int_{\vk} e^{i \vk \cdot \vx} \int_{\vk_1}\int_{\vk_2}\int_{\vq_1}\int_{\vq_2}\int_{\vp_1}\int_{\vp_2} \delta^{\rm D}(\vk+\vk_1+\vk_2)\delta^{\rm D}(\vk_1+\vq_1+\vq_2)\delta^{\rm D}(\vk_2+\vp_1+\vp_2)\big[\\
    &\quad\quad\quad\quad \frac{\vq_2 \cdot \vp_2}{k^2} W(\vq_1)  W(\vp_1)  \td_F(\vq_2) \td_F(\vp_2) 
    \big]\\ \nonumber
    \star 3 &= \int_{\vk} e^{i \vk \cdot \vx} \int_{\vk_1}\int_{\vk_2}\int_{\vq_1}\int_{\vq_2}\int_{\vp_1}\int_{\vp_2} \delta^{\rm D}(\vk+\vk_1+\vk_2)\delta^{\rm D}(\vk_1+\vq_1+\vq_2)\delta^{\rm D}(\vk_2+\vp_1+\vp_2)\big[\\
    &\quad\quad\quad\quad \frac{\vq_1 \cdot \vp_2}{k^2} W(\vq_1)  W(\vp_1)  \td_F(\vq_2) \td_F(\vp_2) 
    \big]\\ \nonumber
    \star 4 &= \int_{\vk} e^{i \vk \cdot \vx} \int_{\vk_1}\int_{\vk_2}\int_{\vq_1}\int_{\vq_2}\int_{\vp_1}\int_{\vp_2} \delta^{\rm D}(\vk+\vk_1+\vk_2)\delta^{\rm D}(\vk_1+\vq_1+\vq_2)\delta^{\rm D}(\vk_2+\vp_1+\vp_2)\big[\\
    &\quad\quad\quad\quad \frac{\vq_2 \cdot \vp_1}{k^2} W(\vq_1)  W(\vp_1)  \td_F(\vq_2) \td_F(\vp_2) 
    \big]
\end{align} 
One can then further simplify and take expectation values of these terms to produce integrals over the flux power spectrum with appropriate derivative coupling kernels modulated by the window factors\footnote{The last two terms are the same for $F_2$ (but of course are not for two distinct fields, e.g. for the shift arising in $\mathcal{S}_7$).}.
However, this last expression explicitly shows how the window function gradients complicate the standard approach to window convolution for power spectra/bispectra when applied to skew spectra via mode mixing.

Similarly, for the tidal term (in $\mathcal{S}_1$ and $\mathcal{S}_3$), and for the LOS-dependent and $SO(2)$-type skew spectra ($\mathcal{S}_{4}-\mathcal{S}_{26}$), derivatives of the window and inverse Laplacian terms have to be computed, which suffer from the same difficulty as above -- we leave this to future work.

\section{Equivalence principle fixing of displacement bias coefficients} \label{app:disp_ep}

In this Appendix, we fix the displacement bias coefficients by performing a calculation similar to that of Ref.~\cite{Fujita:2020_monkey_bias}.
We will follow the notation of Ref.~\cite{Fujita:2020_monkey_bias} here, and rephrase the results in terms of the $SO(2)$ bias expansion coefficients discussed in Section~\ref{sec:derive_2d}.

Rewriting eqn.~\ref{eq:K_kernels} in terms of all mathematically possible quadratic operators consistent with the $SO(2)$ symmetries for the linear and quadratic kernels, we have
\begin{align}
\label{eqn:K_kernels_gen}
    \tilde{K}_1(\vk;\hat{\mathbf{z}},\alpha) &= a_1^{(\alpha)} +a_2^{(\alpha)} \frac{k_{\parallel}^2}{k^2}\\
    \tilde{K}_2(\vk_1,\vk_2;\hat{\mathbf{z}},\alpha) &=b_1^{(\alpha)} + b_2^{(\alpha)} \frac12 \mu_{12}\left(\frac{k_1}{k_2} + \frac{k_2}{k_1}\right) + b_3^{(\alpha)}\mu_{12}^2 \nonumber \\
    &\quad+\frac{k_{3\parallel}^2}{k_3^2}\left[c_1^{(\alpha)} + c_2^{(\alpha)}\frac12 \mu_{12}\left(\frac{k_1}{k_2} + \frac{k_2}{k_1}\right) + c_3^{(\alpha)}\mu_{12}^2\right] \nonumber \\
    &\quad+d_1^{(\alpha)}k_{1\parallel}k_{2\parallel}\left(\frac{1}{k_1^2} + \frac{1}{k_2^2} \right) + d_2^{(\alpha)}\frac{k_{1\parallel}k_{2\parallel}}{k_1^2 k_2^2}\left(k_{1\parallel}^2 + k_{2\parallel}^2 \right)  \nonumber \\
    &\quad+e_1^{(\alpha)}\frac12\left(\mu_1^2 + \mu_2^2 \right) + e_2^{(\alpha)} \mu_1^2 \mu_2^2 \nonumber \\
    &\quad+g_1^{(\alpha)} \mu_{1} \mu_{2} \mu_{12}  \nonumber,
\end{align}
where we neglected constant numerical factors, which do not influence the final results.

The IR-divergent part of the bispectrum with a matter mode in the soft limit then corresponds to the terms associated with: 1. $b_2^{(\alpha)}$ in the isotropic displacement term applied to $\delta$ in $F_{2}$ (as in the real-space case), 2. $c_2^{(\alpha)}$ in the isotropic displacement term in $G_{2}^\parallel$, 3. $d_1^{(\alpha)}$ in the line-of-sight displacement term applied to $\delta$ (e.g., for galaxies displaced from real space, the shift of the argument of $\delta$ due to RSD), 4. $d_2^{(\alpha)}$ in the line-of-sight displacement term applied to $\delta^\parallel$.

The IR-divergent contributions to the (symmetrized) matter-tracer-tracer bispectrum (when the matter mode is soft) are then
\begin{align}
    \label{eqn:IR_div_mtt}
    \frac{\lim_{q\to0}\langle\delta(\vq)\delta_t^{(\alpha)}(\vk_1)\delta_t^{(\beta)}(\vk_2)\rangle'}{\left(1 + f \mu_q^2\right)P_m(q)\langle\delta_t^{(\alpha)}(\vk_1)\delta_t^{(\beta)}(\vk_2)\rangle'} 
    &= \frac{k_1}{q}\bigg[\left(a_{1}^{(\alpha)}+\mu_1^2a_2^{(\alpha)}\right)\left(\mu_{q1}b_2^{(\beta)}+\mu_{q1}\mu_1^2 c_2^{(\beta)} + \mu_1\mu_q d_1^{(\beta)} + \mu_1^3\mu_q d_2^{(\beta)}\right)\\
    &\quad-\left(a_{1}^{(\beta)}+\mu_1^2a_2^{(\beta)}\right)\left(\mu_{q1}b_2^{(\alpha)}+\mu_{q1}\mu_1^2 c_2^{(\alpha)} + \mu_1\mu_q d_1^{(\alpha)} + \mu_1^3\mu_q d_2^{(\alpha)}\right)\bigg]\nonumber,
\end{align}
where, in the squeezed limit, $\mu_{2q}\to-\mu_{q1}$, $\mu_{2}\to-\mu_{1}$.

To satisfy the consistency relation (CR) this squeezed limit set of contributions on the RHS of eqn.~\ref{eqn:IR_div_mtt} must vanish in general (e.g., when $\mu_1 \neq 0$) at equal time. 
Writing this as a polynomial in $\mu_1$, this means all coefficients at each order in $\mu_1^m$ must simultaneously vanish.
Absorbing overall shared factors of each coefficient, we thus obtain the following relationship between bias parameters
\begin{align}
    \label{eqn:tt_mu1s}
    \mu_1^0:& ~ a_1^{(\beta)}b_2^{(\alpha)}=a_1^{(\alpha)}b_2^{(\beta)}\\
    \mu_1^1:& ~ a_2^{(\beta)}c_2^{(\alpha)}=a_2^{(\alpha)}c_2^{(\beta)}\\
    \mu_1^4:& ~ a_1^{(\beta)}d_1^{(\alpha)}=a_1^{(\alpha)}d_1^{(\beta)}\\
    \mu_1^5:& ~ a_2^{(\beta)}d_2^{(\alpha)}=a_2^{(\alpha)}d_2^{(\beta)}
\end{align}
The two other expressions from the terms proportional to $\mu_1^2$ and $\mu_1^3$ are redundant. 
We see that in the $\mu_1=0$ case, we recover the real-space relation fixing the displacement term to the $F_2$ value $a_{1}^{(\alpha)}b_{2}^{(\beta)}=a_{1}^{(\alpha)}b_{2}^{(\alpha)}$, as expected.
Considering also the special case where one of the (hard-mode) tracers is just the matter field, we further obtain
\begin{align}
    \label{eqn:cross_mt_mu1s}
    \mu_1^0:& ~b_{2}^{(\alpha)} = a_{1}^{(\alpha)}\\
    \mu_1^1:& ~c_{2}^{(\alpha)} = a_{2}^{(\alpha)}\\
    \mu_1^4:& ~d_{1}^{(\alpha)} = f ~a_{1}^{(\alpha)}\\
    \mu_1^5:& ~d_{2}^{(\alpha)} = f ~a_{2}^{(\alpha)},
\end{align}
which fixes the values of the bias parameters.

We emphasize that the adiabatic mode condition \cite{horn_consistency} is what ensures the redshift-space CR can be generalized in this way to account for the case of $SO(2)$ tracers beyond the case of LOS-displaced real-space galaxies.
One could also consider the tracer-tracer-tracer bispectrum, but since the soft mode is fixed by the adiabatic condition (eliminating large-scale velocity bias), the soft mode tracer operator will reduce to that of matter, giving back the matter-tracer-tracer case.

From these expressions we can easily see that the  associated bias parameters of real space galaxies that are displaced into redshift space can be recovered by specifying that $a_{2}^{(\alpha)} = f~b_\eta= f$, or, $b_\eta=1$.
Since in this case, e.g. in an SPT model, $b_{2}^{(\alpha)}$ and $c_{2}^{(\alpha)}$ fix the coefficients of the $F_2,G_2^\parallel$ terms in eqn.~\ref{eq:K_kernels}.

\section{Mass \& momentum conservation fixing of bias coefficients under mass/number-conserving displacement} \label{app:consv_Z2} 

In this Appendix, we show that imposing mass/number and momentum conservation on the relevant part of the $K_2$ kernel forces certain relationships between the bias parameters (roughly following the procedure of Ref.~\cite{DAmico:2021_lss_bootstrap}, though see also Ref.~\cite{Chen:2021} for related discussion of the nonlinear PT kernels).
For a fixed value of $b_\eta$, this leads to the usual $Z_2$ kernel for redshift-space galaxies.

Initially, we work with the expression in the basis in terms of all mathematically possible operators under the $SO(2)$ symmetry eqn.~\ref{eqn:K_kernels_gen}, and then map back to the basis of eqn.~\ref{eq:K_kernels}.
With no constraints, eqn.~\ref{eqn:K_kernels_gen} has 11 free parameters, while $Z_2$ has only 3 (those corresponding to real-space galaxies, $b_1, b_2,b_{\mathcal{G}_2}$.
The equivalence principle fixes 4 of these coefficients (see Appendix~\ref{app:disp_ep}), leaving 7 free. 
We now show that mass and momentum conservation fix many of these remaining free bias coefficients.

To impose mass/number conservation on the anisotropic part of $\tilde{K}_2$, we require 
\begin{equation}
    \label{eqn:massnum_consv} 
    \lim_{\vk_2\to-\vk_1} \left(\tilde{K}_2(\vk_1,\vk_2,\hat{z},\alpha) - \tilde{Q}_2(\vk_1,\vk_2,\alpha)\right) \to 0,
\end{equation}
where $\tilde{Q}_2$ is the kernel corresponding to the isotropic part of $\tilde{K}_2$.
This condition removes all constraints on the $b_{n}^{(\alpha)}$ coefficients\footnote{Applying the limit to $\tilde{K}_2$ in its entirety would also enforce that number/mass conservation, which is not satisfied for 3D tracers, but is satisfied by the PT matter field itself}.

This leads to a polynomial in $\mu_{1}^m,\mu_{12}^n$, and requiring the coefficients to vanish gives 
\begin{align}
    \label{eqn:massnum_bias}
    \mu_{1}^{0} \mu_{3}^2&:~ c_1 -c_2 + c_3 = 0,\\
    \mu_{1}^{2} \mu_{3}^0&:~ e_1 -d_1 + g_1 = 0,\\
    \mu_{1}^{4} \mu_{3}^0&:~ e_2 -d_2 = 0,
\end{align}
where we suppress the common tracer index ($\alpha$).
The expression on first line defines a relation for the  contribution related to 2nd-order-in-perturbations kernel (from $\Phi^{(2)}$ ).
The next two lines impose constraints on the terms arising from products of first order terms with line-of-sight dependence.
As previously mentioned, the equivalence principle already fixes $c_2, d_1, d_2$ in terms of the linear bias coefficients $a_1,a_2$ (see Appendix~\ref{app:disp_ep}).

To impose momentum conservation, we require 
\begin{equation}
    \label{eqn:mom_consv} 
    \lim_{\vk_2\to-\vk_1} \left(\frac{\partial}{\partial k_2^i}\left[\tilde{K}_2(\vk_1,\vk_2,\hat{z}) - \tilde{Q}_2(\vk_1,\vk_2,\alpha) \right]\right)\to 0.
\end{equation}

Proceeding similarly to above, we find two terms that must vanish proportional to $k_1^2$ and to $k_{1\parallel}^2$, which requires the following relationships between the bias coefficients
\begin{align}
    \label{eqn:mom_bias}
        k_1^2&:~ d_1 - e_1 = 0\\
        k_{1,\parallel}^2&:~ d_2 - e_2 - g_1 = 0.
\end{align}
This further simplifies the relationship between the coefficients on the last two lines of eqn.~\ref{eqn:massnum_bias}\footnote{Note that we did not actually need to subtract the isotropic part of the kernel $\tilde{Q}_2$ in eqn.~\ref{eqn:mom_consv}, as $\tilde{Q}_2$ already satisfies momentum conservation (even though it does not satisfy number conservation).}.

The combination of the 9 
constraints on the bias parameters from the equivalence principle (4 from eqns.~\ref{eqn:cross_mt_mu1s}), mass/number conservation (3 
from eqns.~\ref{eqn:massnum_bias}), and momentum conservation (2 from eqns.~\ref{eqn:mom_bias}) can be written as a (underconstrained) linear system.
The resulting matrix for the 12 
bias coefficient variables has rank 8, 
and has the (5D) 
(in terms of 
$a_1,a_2,b_1,b_3,c_1$
)
\begin{align}
    \label{eqn:bias_lin_soln}
    b_2 &= a_1 \\
    c_2 &= a_2\\
    c_3 &= a_2-c_1\\
    d_1 &= f a_1\\
    d_2 &= f a_2\\
    e_1 &= f a_1\\
    e_2 &= f a_2\\
    g_1 &= 0
\end{align}

Written in the usual basis (of eqn.~\ref{eq:K_kernels}) this means $b_1,b_\eta,b_2,b_{\mathcal{G}_2}$ can be taken as the free parameters.
This nearly reduces to the case of $Z_2$, which additionally requires fixing $b_\eta=1$, or rather fixing the value of $a_2 = f$ (see discussion in Section~\ref{subsubsec:ep}).

\bibliographystyle{aux_files/JHEP}
\bibliography{references}

\end{document}